\documentclass[usegraphicx,usenatbib,useAMS]{mn2e}
\usepackage{times}
\newcommand{\be}{\begin{equation}}
\newcommand{\ba}{\begin{eqnarray}}
\newcommand{\ee}{\end{equation}}
\newcommand{\ea}{\end{eqnarray}}

\newcommand{\url}{\tt}% 
\def\lesssim{\mathrel{\hbox{\rlap{\hbox{\lower4pt\hbox{$\sim$}}}\hbox{$<$}}}}
\def\gtrsim{\mathrel{\hbox{\rlap{\hbox{\lower4pt\hbox{$\sim$}}}\hbox{$>$}}}}
\def\apj{ApJ}
\def\apjl{ApJL}
\def\apjs{ApJS}
\def\aj{AJ}
\def\mnras{MNRAS}

\def\aap{{\em A.\&A}}
\def\gtsima{$\; \buildrel > \over \sim \;$}
\def\ltsima{$\; \buildrel < \over \sim \;$}
\def\gsim{\lower.5ex\hbox{\gtsima}}
\def\lsim{\lower.5ex\hbox{\ltsima}}
\def\simgt{\lower.5ex\hbox{\gtsima}}
\def\simlt{\lower.5ex\hbox{\ltsima}}
\def\simpr{\lower.5ex\hbox{\prosima}}

\def\simless{\mathbin{\lower 3pt\hbox
   {$\rlap{\raise 5pt\hbox{$\char'074$}}\mathchar''7218$}}}   % < or of order
\def\simgreat{\mathbin{\lower 3pt\hbox
   {$\rlap{\raise 5pt\hbox{$\char'076$}}\mathchar''7218$}}}   % > or of order
\def\apj{ApJ}
\def\apjs{ApJS}
\def\apjl{ApJL}
\def\aap{A\&A}
\def\aj{AJ}
\def\mnras{MNRAS}

\begin{document}

\title[Simulating Cosmic Reionization at Large Scales]{Simulating Cosmic 
Reionization at Large Scales II: the 21-cm Emission Features and Statistical 
Signals}
\author[G. Mellema, et al.]{Garrelt Mellema
$^{1,2}$\thanks{Current address: Stockholm Observatory, AlbaNova University Centre, SE-106 91 Stockholm, Sweden; e-mail: garrelt@astro.su.se},
Ilian T. Iliev$^{3}$,
Ue-Li~Pen$^3$,
Paul~R.~Shapiro$^4$
\\
$^1$ ASTRON, P.O. Box 1, NL-7990 AA Dwingeloo, The
  Netherlands \\
$^2$ Sterrewacht Leiden, P.O. Box 9513, NL-2300 RA Leiden, The
  Netherlands \\
$^3$ Canadian Institute for Theoretical Astrophysics, University
  of Toronto, 60 St. George Street, Toronto, ON M5S 3H8, Canada\\
$^4$ Department of Astronomy, University of Texas,
  Austin, TX 78712-1083, USA
}
%\date{DRAFT \today\, DRAFT}
\pubyear{2006} \volume{000} \pagerange{1}
%\twocolumn
\maketitle\label{firstpage}

\begin{abstract}
  We present detailed predictions for the redshifted 21~cm signal from the
  epoch of reionization. These predictions are obtained from radiative
  transfer calculations on the results of large scale (100/$h$~Mpc), high
  dynamic range, cosmological simulations. We consider several scenarios for
  the reionization history, of both early and extended reionization. From the
  simulations we construct and analyze a range of observational
  characteristics, from the global signal, via detailed images and spectra, to
  statistical representations of rms fluctuations, angular power spectra, and
  probability distribution functions to characterize the non-Gaussianity of the
  21~cm signal. We find that the different reionization scenarios produce quite
  similar observational signatures, mostly differing in the redshifts of 50\%
  reionization, and of final overlap. All scenarios show a gradual transition
  in the global signatures of mean signal and rms fluctuations, which would
  make these more difficult to observe. Individual features such as deep gaps
  and bright peaks are substantially different from the mean and mapping these
  with several arcminutes and 100s of kHz resolution would provide a direct
  measurement of the underlying density field and the geometry of the
  cosmological HII regions, although significantly modified by peculiar
  velocity distortions. The presence of late emission peaks suggest these
  to be a useful target for observations. The power spectra during reionization
  are strongly boosted compared to the underlying density fluctuations. The
  strongest statistical signal is found around the time of 50\% reionization
  and displays a clear maximum at an angular scale of $\ell\sim 3000$--$5000$.
  We find the distribution function of emission features to be strongly
  non-Gaussian, with an order of magnitude higher probability of bright
  emission features. These results suggest that observationally it may be
  easier to find individual bright features than deriving the power spectra, 
  which in its turn is easier than observing individual images.

\end{abstract}


\section{Introduction}

Observations of the redshifted 21-cm line of hydrogen are currently
emerging as the most promising approach for the direct detection of the
epoch of reionization, and possibly even of the preceding period, the
Cosmic Dark Ages. Several large radio interferometer arrays are
currently either operational, under construction or being planned that
have the potential to detect this redshifted 21-cm emission.  These
projects include
PAST\footnote{\url{http://web.phys.cmu.edu/$\sim$past/}} (already
operating in north-western China),
GMRT\footnote{\url{http://www.ncra.tifr.res.in}} (operating in India),
LOFAR\footnote{\url{http://www.lofar.org}} (under construction in the
Netherlands),
MWA\footnote{\url{http://web.haystack.mit.edu/arrays/MWA}} (to be
located in Western Australia), and
SKA\footnote{\url{http://www.skatelescope.org}}.

The observations will be complicated due to the combined effects of (in order
of increasing distance from us) possible radio frequency interference from
terrestrial emitters (in particular in the 87-108~MHz FM band), ionospheric
fluctuations, the galactic foreground, and unresolved intergalactic radio
sources, all of which are much stronger than the expected redshifted 21-cm
signal. Therefore it is crucial to understand the characteristics of the
reionization 21-cm signal in some detail. This can help both in the planning of
the experiments, so that they are optimized for the expected signal, and 
once data is available, can direct us in separating the strong foregrounds 
and interpreting the signal correctly. Predicting the unique signatures of
reionization is quite hard. The $\Lambda$CDM framework for cosmological
structure formation is now well-established, the fundamental cosmological
parameters are increasingly better constrained, and the basic physical
processes during reionization are fairly well-understood. However, we will be
probing redshifts for which little or no observational data is currently
available, and many of the relevant parameters are at best 
poorly constrained. For example, a number of choices remain open regarding the
nature of the ionizing sources during reionization, their numbers, photon
production efficiencies and spectra.

Another complication is the large dynamic range required from any
simulation which aims to predict the redshifted 21-cm emission. Within
the hierarchical structure formation paradigm the ionizing photon
emissivity during reionization is dominated by numerous dwarf-size
galaxies, rather than by larger ones, which at high redshifts are too
rare to make an appreciable contribution. On the other hand, the
strong source clustering at high redshifts means that ionized bubbles
quickly overlap locally and each H~II region typically contains a
large number of sources. This leads to the formation of large ionized
regions, of size tens of comoving Mpc or more, which requires
simulation volumes of size $\sim100$~Mpc for proper treatment.
Cosmological simulations resolving dwarf galaxies in such large
volumes are only now becoming possible. Even more difficult and
computationally-intensive is the transfer of ionizing radiation from
the tens of thousands up to millions of individual galaxies found in
such a large volume. {A further challenge is the absorption of
photons by collapsed objects, which happens at even smaller
scales. This however, only becomes important after or around final
overlap, when the opacity of the IGM to ionizing photons has dropped
to a few percent and these (proto-)galaxies (e.g.~Lyman Limit Systems)
become the main source of opacity for ionizing photons.}

To date there have been only a few cosmological simulations which studied the
redshifted 21-cm. Most of these resolved this difficulty by considering small
enough regions (thus achieving the high resolution required) either during the
Cosmic Dark Ages \citep{2005astro.ph.12516S,2006ApJ...637L...1K} or during
reionization \citep{2002ApJ...577...22C,2003ApJ...596....1C,
  2004ApJ...608..611G,2004MNRAS.347..187F,2006MNRAS.369L..66V}. An alternative
approach is to simulate large regions at low resolution and
thus without resolving individual small sources, and to include their effect
in an averaged, approximate way \citep{2005ApJ...633..552K}.

Recently we presented the first N-body and radiative transfer simulations
which considered a sufficiently large volume, and at the same time resolved all
dwarf galaxies in that volume and accounted for their individual contributions
to reionization \citep[][hereafter Paper I]{topologypaper}.  We first
performed an N-body simulation with 4.3 billion particles in a
(100$\,h^{-1}$~Mpc)$^3$ volume. This provided us with all halos with masses
above $2.5\times10^9M_\odot$ and the cosmological evolution of the density
field. We then imported these results into a new fast and precise radiative
transfer code called $C^2$-Ray \citep{methodpaper}. In Paper~I we presented
our simulation methodology, as well as the results for the large-scale
geometry (often also referred to as topology) of reionization, i.e.\ the size
and number distributions of the H~II regions in space, their evolution in
time, and the power spectra of the resulting neutral and ionized density
fields. We showed that the fluctuations in both the neutral and the ionized
density are strongly boosted due to the patchiness of reionization. We also
demonstrated that the probability distribution function (PDF) of the ionized
fraction and the ionized density are generally strongly non-Gaussian at all
scales and quantified their level of departure from non-Gaussianity for the
first time.  We also derived the PDF and the mean optical depths to
Ly-$\alpha$ photons (related to the Gunn-Peterson effect in spectra of distant
sources). In this paper we present detailed predictions for the redshifted
21-cm signal, including results from several new simulations in addition to
the original one from Paper~I. We focus on generic predictions of the
observational properties from our simulations, rather than aiming our analysis
to specific radio-interferometry arrays.

The lay-out of this paper is as follows. In Sect.~\ref{signal_sect} we
describe the basic procedures of extracting the 21-cm signal from the data and
the assumptions that go into that. In Sect.~\ref{sim_sect} we discuss our
simulations and their basic parameters and features. In
Sect.~\ref{mean_bg_sect} we present the evolution of the mean 21-cm signal and
its implications. Next, in Sect.~\ref{reion_geom_sect} we discuss how the
reionization geometry would be seen at the redshifted 21-cm line and the
dependence of this on the adopted observational beam shape. The first 21-cm
signals to be detected would quite possibly be individual, bright features.
These are discussed in Sect.~\ref{indiv_sect}. Finally, the statistics of the
21-cm emission signal is discussed in Sect.~\ref{stat_sect}. Our conclusions
are summarized in Sect.~\ref{summary_sect}.

Throughout this paper we use the concordance flat ($\Omega_k=0$)
$\Lambda$CDM cosmology with parameters
($\Omega_m,\Omega_\Lambda,\Omega_b,h,\sigma_8,n)=(0.27,0.73,0.044,0.7,0.9,1)$
\citep[based on the first year WMAP data][]{2003ApJS..148..175S},
where $\Omega_m$, $\Omega_\Lambda$, and $\Omega_b$ are the total
matter, vacuum, and baryonic densities in units of the critical
density, $\sigma_8$ is the standard deviation of linear density
fluctuations at present on the scale of $8 h^{-1}{\rm Mpc}$, and $n$
is the index of the primordial power spectrum of density fluctuations.

\section{The redshifted 21-cm signal}
\label{signal_sect}
The 21-cm radio line, in either emission or absorption is due to a spin-flip 
transition of neutral hydrogen atoms. This transition quickly enters 
into equilibrium with the CMB photons and hence it needs to be decoupled 
from the CMB by some mechanism  in order to become observable. This could 
occur either through collisions with other hydrogen atoms and free electrons
\citep{1956ApJ...124..542P,1959ApJ...129..536F,1969ApJ...158..423A,
2005ApJ...622.1356Z} 
or through Ly-$\alpha$ photon pumping
\citep{1952AJ.....57R..31W,1959ApJ...129..536F,2006MNRAS.367..259H,
2005astro.ph.12206C}. 

The differential brightness temperature with respect to the CMB of 
the redshifted 21-cm emission is determined by the density of neutral 
hydrogen, $\rho_{\rm HI}$, and its spin temperature, $T_{\rm s}$ \citep[see
e.g.][for detailed discussions]{2004ApJ...615....7M,2005astro.ph.12516S}. 
It is given by 
\begin{equation}
  \delta T_b =\frac{T_{\rm S} - T_{\rm CMB}}{1+z}(1-e^{-\tau})
\label{temp21cm}
\end{equation}
where $z$ is the cosmological redshift, $T_{\rm CMB}$ is the temperature of 
the CMB radiation at redshift $z$, and the optical depth $\tau$ is 
\citep[e.g.][]{2002ApJ...572L.123I}:
\ba
\tau(z)&=&\frac{3\lambda_0^3A_{10}T_*n_{HI}(z)}{32\pi T_S H(z)}=
\frac{0.28}{T_S}\left(\frac{1+z}{10}\right)^{3/2}(1+\delta) 
\label{tau}
\ea
where $\lambda_0=21.16$~cm is the rest-frame wavelength of the line, 
$A_{10}=2.85\times10^{-15}\,\rm s^{-1}$ is the Einstein A-coefficient,
$T_*=0.068$~K corresponds to the energy difference between the two levels, 
$1+\delta={\rho_{\rm HI}}/{ \langle \rho_H \rangle}$ is the mean density 
of neutral hydrogen in units of the mean density of hydrogen at redshift 
$z$, and $H(z)$ is the redshift-dependent Hubble constant,
\begin{eqnarray}
  H(z)&=&H_0E(z)\nonumber\\
  &\equiv& H_0[\Omega_{\rm m}(1+z)^3+\Omega_{\rm k}(1+z)^2+
    \Omega_\Lambda]^{1/2}\\ 
  &\approx&H_0\Omega_{\rm m}^{1/2}(1+z)^{3/2},\nonumber
\end{eqnarray}
{where $E(z)$ is the redshift dependent part, as defined here,}
$H_0$ is the value at the present day, and the last expression is
valid for $z\gg 1$.

Assuming $\tau \ll 1$ (which is always correct for the redshifts of interest 
here, except for some lines of sight through mini-halos
\citep[][]{2002ApJ...572L.123I}, equation~\ref{temp21cm} becomes 
\ba 
\delta T_b&\approx& ({\rm 3.1 mK})
h^{-1} \frac{(1+z)^2}{E(z)} \frac{(T_{\rm S} - T_{\rm CMB})}{T_{\rm
S}}(1+\delta)\\
&\approx& ({\rm 27 mK}) \left(\frac{1+z}{10}\right)^{1/2}
  \frac{(T_{\rm S} - T_{\rm CMB})}{T_{\rm S}}
(1+\delta)\,,\nonumber
  \label{temp21cm_approx}
\ea
with $h$ being $H_0$ in units of 100~km~s$^{-1}$~Mpc$^{-1}$.

To obtain an appreciable signal, the spin temperature $T_{\rm S}$ should
differ significantly from $T_{\rm CMB}$.  The gas kinetic temperature itself
is expected to be above the CMB temperature due to heating by shocks, X-rays,
and Ly-$\alpha$ photons, thus the redshifted 21-cm is generally in emission.
Decoupling through collisions is efficient only at very high-z and in 
significantly overdense and heated regions, mostly inside collapsed 
halos \citep{2002ApJ...572L.123I,2003MNRAS.341...81I,2005astro.ph.12516S}. 
Once the first stars turn on, the Ly$\alpha$ photons produced by them can
easily pump the population to the gas kinetic temperature 
\citep[e.g.,][]{2003ApJ...596....1C}. Thus, during reionization one can
assume that globally $T_{\rm S} \approx T_{\rm gas} \gg T_{\rm
CMB}$. In this case the precise value of $T_{\rm S}$ becomes irrelevant, 
and the observable differential brightness temperature is only a function 
of redshift and H~I density. This is the assumption we adopt throughout
this paper. One should note that during the earliest stages of reionization, 
when sources were few and could only affect their local environments
this assumption might not hold \citep{2005ApJ...626....1B,2005astro.ph.12206C}, 
but such effects are still difficult to implement in simulations.

\section{The Simulations}
\label{sim_sect}

\begin{figure}
\begin{center}
\includegraphics[width=3.2in]{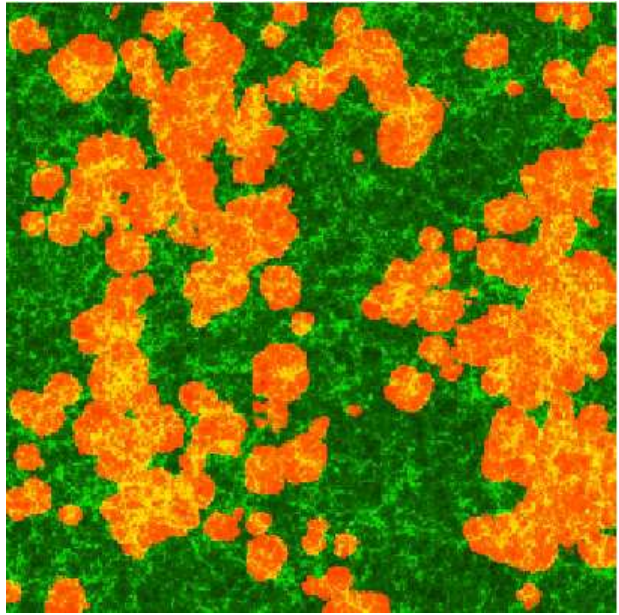}
\caption{A cut through the simulation box of our high resolution, f2000\_406 
run, at z=13.89. The neutral gas density is shown in green. The orange overlay
shows the ionized regions. Each side of the square corresponds to a comoving
length of 100$\,h^{-1}$~Mpc.
\label{f2000_406}}
\end{center}
\end{figure}

The details of our methodology were described in Paper~I. Here we only
summarize the relevant parameters and briefly discuss the main features of
each of the simulations used in the current work. Our base N-body simulation
was performed using the particle-mesh code PMFAST \citep{2005NewA...10..393M}
and has a total of $1624^3=4.3$ billion particles on a $3248^3$ computational
mesh. The computational volume is (100$\,h^{-1}$~Mpc)$^3$. We found all halos
with masses above $2.5\times10^9M_\odot$ (corresponding to 100 particles or
more) on a large number of density time-slices. We then imported these results
into a new fast and precise radiative transfer code called $C^2$-Ray which we
have developed \citep{methodpaper}. The code has been tested in detail against
available analytical solutions \citep{methodpaper} and in comparison with
other radiative transfer codes \citep{2006astro.ph..3199I}. The radiative
transfer runs have mesh resolutions of 203$^3$ and 406$^3$, since the full
N-body mesh, at $3248^3$, is still well beyond the current computational
capabilities. Our ionizing sources correspond to all the halos found in our
volume at the full resolution of the underlying N-body simulation. No
artificial grouping of sources has been employed, with the exception of
combining sources which happen to be inside the same radiative transfer cell,
something that affects only a small fraction of the sources.

We present the results of five simulations, all of which use the same box
size, density fields and halo catalogues, but each with different the sub-grid
physics, as follows. Four of these simulations are run at the same $203^3$
mesh resolution, but making different assumptions about the gas clumping at
small scales and about the photon production efficiencies of the ionizing
sources. The fifth simulation is one run at the higher mesh resolution of
$406^3$, but otherwise adopting the same assumptions.  All simulations and
their basic parameters and features are summarized in Table~\ref{summary}.

\begin{table}
\caption{Simulation parameters and global reionization history results}
\label{summary}
\begin{center}
\begin{tabular}{@{}llllll}
                  & f2000   & f2000\_406 & f250    & f2000C  & f250C\\ [2mm]
\hline\\
mesh              & $203^3$ & $406^3$    & $203^3$ & $203^3$ & $203^3$\\[2mm]
$f_\gamma$        & 2000    & 2000       & 250     & 2000    & 250 \\[2mm]
$C_{\rm subgrid}$ & 1       & 1          & 1       &  $C(z)$ &  $C(z)$\\[2mm]
\hline\\
$z_{50\%}$        & 13.6     & 13.5        & 11.7  & 12.6    & 11.0  \\[2mm] 
$z_{\rm overlap}$ & 11.3     & $\sim 11$   & 9.3   & 10.15   & 8.2\\[2mm] 
$\tau_{\rm es}$   & 0.145    & $\sim 0.14$ & 0.121 & 0.135   & 0.107 \\[2mm]
\hline\\
\end{tabular}
\end{center}
\end{table}

Our first case, which we will call f2000 hereafter, is the simulation we
presented in detail in Paper~I.  It assumes a source photon 
production efficiency (which is a combination of the total number of ionizing
photons emitted by the stars per unit time, the star formation efficiency and 
the escape fraction) of $f_\gamma=2000$ photons per halo atom, which
corresponds to a top-heavy initial mass function (IMF). 
It reaches overlap (defined as more than 99\% ionization by mass) at a
redshift of $z=11.3$. The resulting optical depth for electron scattering is
$\tau_{\rm es}=0.145$, within the 2-$\sigma$ limits of the new WMAP
value, $\tau_{\rm es}=0.09\pm0.03$ \citep{Page}. Our second case is the same
simulation as f2000 but at a higher mesh resolution of $406^3$ (hereafter
called f2000\_406).
In general the results closely match the f2000 run. The density field is
resolved better and the H~II regions have somewhat less spherical shapes than
in the lower resolution run (as could be expected). The global evolution is
slightly slower due to the better-resolved density field, resulting in higher
effective gas clumping. As an illustration we show in Fig.~\ref{f2000_406} a
slice through the simulation volume of the density and ionization structures
at redshift $z=13.89$, extracted from this high-resolution run.

Our third simulation (labelled f250) adopts a lower photon production efficiency
of 250 photons per halo atom, corresponding to either a slightly top-heavy
IMF, or a Salpeter IMF combined with a bit higher escape fraction and star 
formation efficiency at these high redshifts. This simulation reaches overlap 
at redshift $z=9.3$. The resulting optical depth for electron scattering is 
$\tau_{\rm es}=0.121$, within the 1-$\sigma$ limit of the new WMAP value.

The fourth and fifth cases included in this paper, called f2000C and f250C, assume the same source
efficiencies as f2000 and f250, but add the effect of sub-grid gas inhomogeneities, 
described by a mean volume-averaged clumping factor, 
$C_{\rm subgrid}=\langle n^2\rangle/\langle n\rangle^2$, given by 
\be
C_{\rm subgrid}(z)=27.466 e^{-0.114z+0.001328\,z^2}.
\label{clumpfact_fit}
\ee This fit to the small-scale clumping factor is an improved version
of the one we presented in \citet{2005ApJ...624..491I}. To derive it
we used another PMFAST simulation, with the same computational mesh,
$3248^3$, and number of particles, $1624^3$, but a much smaller
computational volume, $(3.5\,\rm h^{-1}~Mpc)^3$, and thus much higher
resolution. These parameters correspond to particle mass of
$10^3M_\odot$ and minimum resolved halo mass of $10^5M_\odot$. This
box size was chosen so as to resolve the scales most relevant to the
clumping - on smaller scales the gas would be Jeans smoothed, while on
larger scales the density fluctuations are already present in our
computational density fields and should not be included again. The
expression in equation~(\ref{clumpfact_fit}) excludes the matter
residing inside collapsed halos since these contribute to the
recombination rate differently from the unshielded IGM. The minihalos
are self-shielded, which results in their lower contribution to the
total number of recombinations than one would infer from simple gas
clumping argument \citep{2004MNRAS.348..753S,2005MNRAS...361..405I},
while the larger halos are ionizing sources and their recombinations
are implicitly included in the photon production efficiency $f_\gamma$
through their escape fraction.  The f2000C case reaches overlap at a
redshift of 10.15, and the case f250C at 8.2. The resulting integral
optical depths for electron scattering are $\tau_{\rm es}=0.135$ and
0.107.

\begin{figure}
\begin{center}
\includegraphics[width=3.5in]{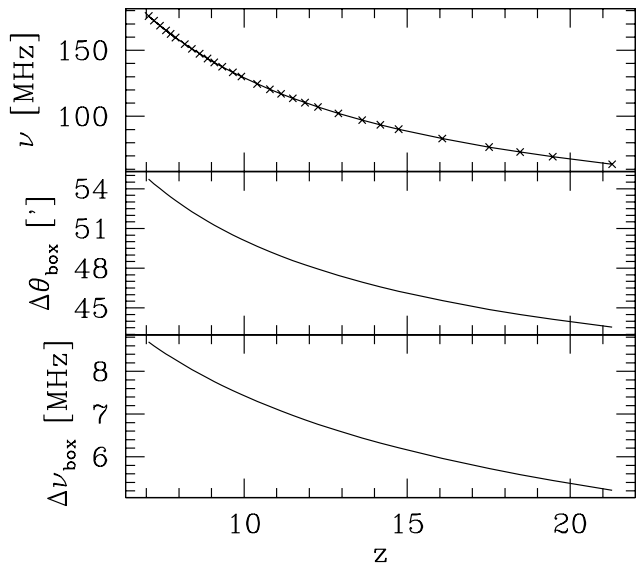}
\caption{Simulation size and output times: (top) the frequency of 
the 21-cm signal (the crosses indicate the redshifts at which we have model
 data), (middle) the angular size on the sky of our computational volume, 
and (bottom) the frequency width of our computational volume, all plotted vs. 
redshift $z$.
\label{box_sizes}}
\end{center}
\end{figure}

The value for the electron scattering optical depth reported from the
3-year WMAP data, $0.09\pm0.03$ \citep{Page}, is substantially below
the first year value
\citep[$0.17\pm0.04$,][]{2003ApJS..148..161K}. Our simulation results
fall roughly within the overlap region between these two results and
within 2-$\sigma$ from the new result. Moderate changes in the
simulation parameters, e.g.\ assuming lower luminosity sources and/or
accounting better for the small-scale gas clumping as in f2000C and
f250C easily extends the reionization process until significantly
later times. However, the value of the optical depth should not be
considered separately from the values of the underlying cosmological
parameters \citep{WMAP3}. \citet{2006ApJ...644L.101A} showed that to first
approximation simulations employing the first year WMAP cosmological
parameters and achieving a $\tau$ of 0.17, would for the 3-year WMAP
cosmological parametes achieve a $\tau$ of 0.1.Based on these results,
we expect that our current results would all be consistent with the
3-year WMAP data. This could be checked with future simulations.

In this paper we will use results from simulations, f2000, f2000C,
f250 and f250C, and results from f2000\_406 only as comparison to
f2000. A more detailed comparison of their different reionization
histories and implications will be the subject of a future paper.

\begin{figure}
\begin{center}
  \includegraphics[width=3.5in]{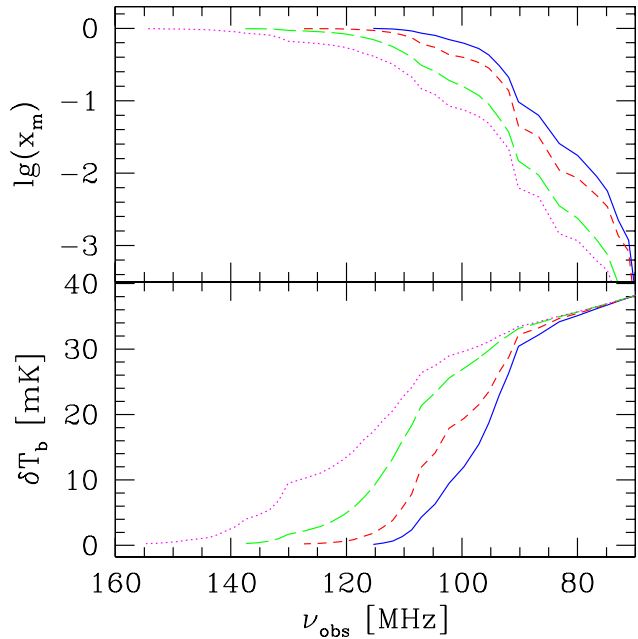}
  \caption{Evolution of (top) the mean mass ionized fraction, $x_m$, and
    (bottom) the mean differential brightness temperature, $\delta T_b$,
    both as a functions of frequency for f2000 (solid, blue), f2000C
    (short-dashed, red), f250 (long-dashed, green), f250C (dotted,
    magenta).
    \label{meantbzfig}}
\end{center}
\end{figure}

For reference we show in Fig.~\ref{box_sizes} the frequency of the 
redshifted 21-cm emission, $\nu_z$ (top), the angular size on the sky, 
$\Delta\theta_{\rm box}$ (middle), and its total bandwidth, 
$\Delta \nu_{\rm box}$(bottom), versus redshift, for our 
$100\,h^{-1}\rm Mpc^3$ comoving computational volume. These quantities are
given by
\begin{eqnarray}
  \nu_z & = & \nu_0 / (1+z)\\
  \Delta\theta_{\rm box} & = & \frac{L_{\rm box}}{(1+z)D_A(z)}\\
  \Delta \nu_{\rm box} & = & \frac{\nu_0 H_0 E(z) L_{\rm box}}{c(1+z)^2}\,,
\end{eqnarray}
where $D_A(z)$ is the angular diameter distance, $L_{\rm box}=100/h\,\rm Mpc$
is the comoving size of our box, $c$ is the speed of light and $\nu_0
=1.420$~GHz is the rest-frame frequency of the line. We see that our
computational box corresponds to almost one degree on the sky and a range of
5.5 to 7 MHz in frequency. The intrinsic resolution of our data is
approximately $14\arcsec$ ($7\arcsec$) in the spatial direction, and 30~KHz
(15~kHz) in frequency for our $203^3$ ($406^3$) meshes. None of
the currently planned experiments will achieve a similar spatial resolution. The
estimates for both PAST and LOFAR are that they will have synthesized beams of
$\sim3$\arcmin.  Their intrinsic frequency resolution will actually be somewhat
higher than in our simulations, e.g. around 1~kHz for LOFAR. However, to
obtain sufficient flux sensitivity, the data will be integrated over
significantly larger bandwidths, e.g. around 200~kHz for LOFAR, which
approximately corresponds to 7 (14) of our cells at $203^3$ ($406^3$)
resolution.

\section{The evolution of the mean background}
\label{mean_bg_sect}

The simplest global measure of the progress of reionization is the
evolution of the globally-integrated mean 21-cm signal. In
Figure~\ref{meantbzfig} we show how the mass-weighted ionized fraction
$x_m$ and the mean differential brightness temperature, $\delta T_b$
evolve against observed frequency, $\nu_{\rm obs}$ for f2000, f2000C,
f250 and f250C. The first thing to note is that the reionization is
significantly delayed, by $\Delta z\sim1$ compared to f2000, when the
small-scale clumping of the gas is included, and even more so, by
$\Delta z\sim2$, when the ionizing sources are less efficient photon
producers. Accordingly, the mean signal disappears and becomes
undetectable at frequencies above $\sim110$~MHz in f2000, but does so
at much higher frequencies, {$\sim120$~MHz ($\sim135$~MHz,
$\sim145$~MHz) in f2000C (f250, f250C).} At face value this seems to imply
that our f2000 reionization history will be impossible to observe in
places with high FM interference, as is the case with e.g.\ LOFAR and
GMRT, which operate above 115~MHz, but as we will show below, this is
probably a too simplistic conclusion.

{Note that in the real universe the signal will never go
completely to zero since there will be neutral hydrogen contained in
collapsed objects such as damped Ly$\alpha$ systems and Lyman limit
systems. However, at the redshifts we are considering these systems
will never contain more than a few tenths of a percent of the mass in our
computational volume, and their combined signal will be too weak to be
observable at 21~cm.}
\begin{figure*}
\includegraphics[width=7.2in]{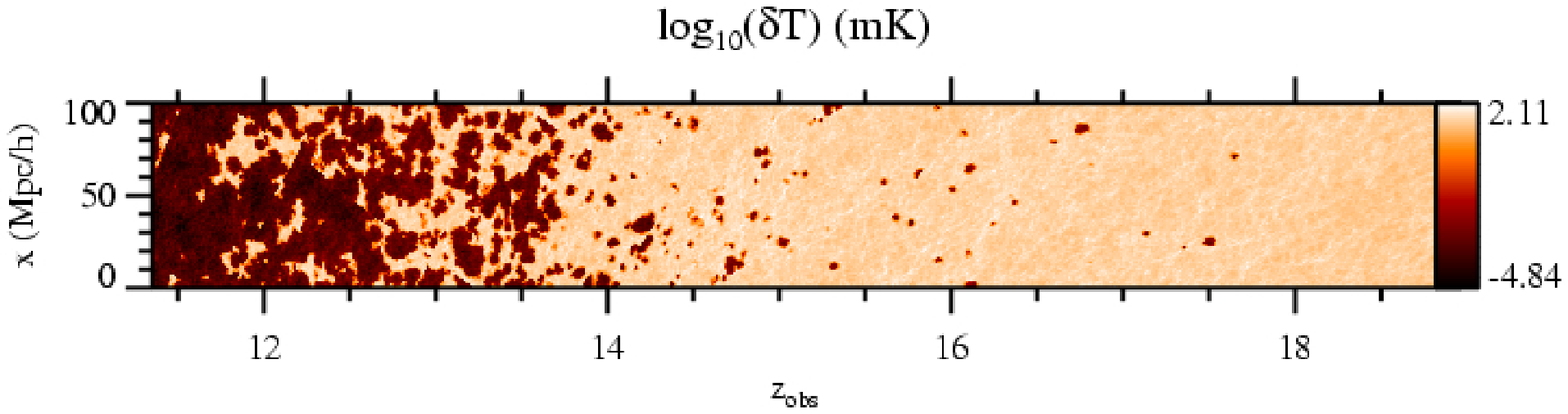}
%\vskip -2cm
\includegraphics[width=7.2in]{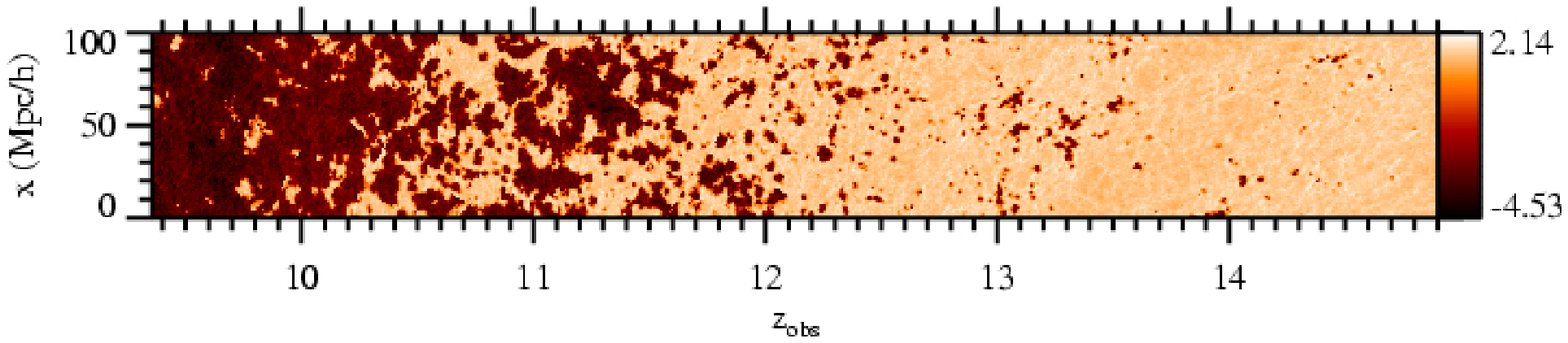}
\caption{Position-redshift slices from the f2000 (top) and f250 (bottom) 
  simulation. The spatial scale is in comoving units. These slices illustrate
  the large-scale geometry of reionization seen at 21-cm and the significant
  local variations in reionization history. Observationally they correspond to
  slices through an image-frequency volume.
\label{pencil}}
\end{figure*}

It has been proposed that one could possibly detect the transition from
fully-neutral gas to almost fully-ionized as a ``global reionization step''
over the whole sky \citep{1999A&A...345..380S}. To be detectable against the
smoothly varying foregrounds, such a transition should occur fairly quickly,
resulting in a sharp drop in the global 21-cm signal in frequency space when
observed with a single dish radio telescope. All of our models show a rather
gradual transition, with the mean signal decreasing by $\sim20$~mK over
$\sim20$~MHz.  Detecting such a transition is in principle well within the
capabilities of even a 10~m radio dish \citep{1999A&A...345..380S}. However, a
gradual change over 20 to 30~MHz would be difficult to disentangle from the
strong foregrounds. The practicality of such observation would thus depend
strongly on the (still highly-uncertain) detailed properties of the spectral
index variations of the foregrounds and how well one can model and subtract
them.

\section{Reionization geometry seen in redshifted 21-cm emission}
\label{reion_geom_sect}

\subsection{Global evolution}
\label{global_evol_sect}

We start with an overview of the progress of reionization in our simulations
f2000 and f250, as seen in the 21-cm emission line. We present a visualization
of the evolution of the differential brightness temperature in
Fig.~\ref{pencil}. We constructed these slices by linearly interpolating the
evolution of the whole simulation box between our discrete time outputs,
wrapping the box periodically after each complete crossing of the volume, and
taking slices through the resulting long box. The vertical axis corresponds to
the complete extend of our simulation volume of $100\,h^{-1}$~Mpc comoving,
while the horizontal axis shows the redshift at that position. Since redshift
corresponds to frequency, these slices are equivalent to slices through an
idealized observational image-frequency volume. We took the slices at an angle
to the simulation cube axes, to avoid artificial periodicity effects 
caused by repeatedly passing through the same structures at later times.  We 
also include the effect of peculiar bulk velocities from our N-body 
simulations, which result in redshift distortions along the line-of-sight (see
Sect.~\ref{LOSspectra} for more discussion on this last point).

Both slices show all the basic evolution trends we discussed in Paper~I. The
reionization starts around $z\sim20$, with a few isolated and highly clustered
ionized regions. Most of the gas is still neutral and quite bright at 21-cm
emission, with some pixels reaching $\delta T_b$ of well over 100~mK. The
Cosmic Web is already well-developed and shows in the images even at these
fairly large scales. For f2000 (Figure~\ref{pencil}, top image) the
first H~II regions quickly merge with each other locally, forming
significantly larger ones, of size $\sim10$ Mpc comoving, already by
$z\sim14$, while at the same time many new ionizing sources emerge and start
expanding their own local H~II regions. By redshifts $z\sim12.5$ many of the
larger H~II regions merge together to form a couple, and later on just a
single, very large and topologically-connected (but highly-nonspherical)
volume of $>10^4\,\rm (Mpc)^3$. Essentially all remaining ionized regions
merge with this large one around $z\sim11.5$ and the last pockets of neutral
gas (mostly in the large voids, which have no local ionizing sources and are
fairly far away from any source) are ionized by $z\sim11$. In the f250 case
(Figure~\ref{pencil}, bottom image) the evolution is a bit slower
due to the weaker sources, and the H~II regions initially smaller, on average.
Essentially the same main evolutionary trends remain in force, with
significant local overlap of H~II regions starting at $z\sim 12$, and the
wide-spread merging of these between redshift $z=11$ and 11.5. These finally
merge into one huge ionized region between $z=10$ and 9.5, although
significant local pockets of neutral gas remain until the end.

While these features are generic and thus seen in all slices, there are a
number of special features to note. Reionization is a highly inhomogeneous
process, with large variation in its progress between different 2D slices of
the same simulation, and even more between different lines-of-sight (LOS).
For example, in the f2000 slice some LOS contain only few 21-cm features after
$z\sim13$, while others go through neutral pockets even at global overlap 
($z\sim 11$). In the f250 case similar LOS behaviour can be seen.  This strong
patchiness until late times will have important implications for direct
observations of high-z Ly-$\alpha$ sources, a point we will address in future
work.

\begin{figure}
\includegraphics[width=3in]{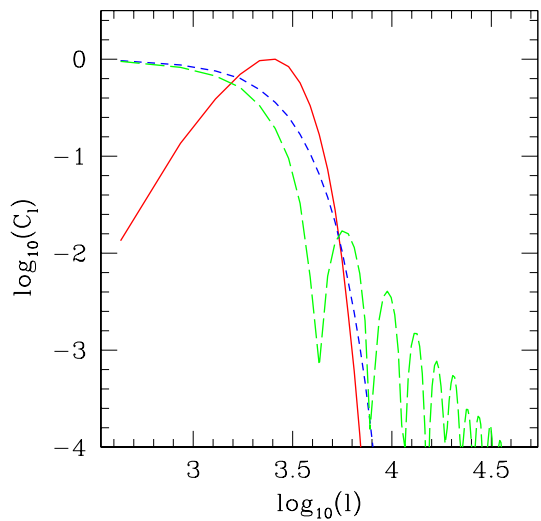}
\caption{Angular power spectra of the compensated Gaussian (red/solid), Gaussian
  (short-dashed/blue), and tophat (long-dashed/green) beams all with FWHM of 3 
  arcmin and each normalized to maximum of 1. The sharp cut-off
  of the tophat beam produces high $\ell$ oscillations, showing that it should
  not be used for Fourier analysis of results. The compensated Gaussian,
  which is the one most closely resembling real-life interferometer beams has
  less power at low $\ell$'s than the Gaussian, reflecting the relative
  insensitivity of compact interferometers for large scale structures.
\label{beams_ps}}
\end{figure}

\subsection{Sky maps}
\label{maps_sect}

\begin{figure*}
\begin{center}
\includegraphics[width=2in]{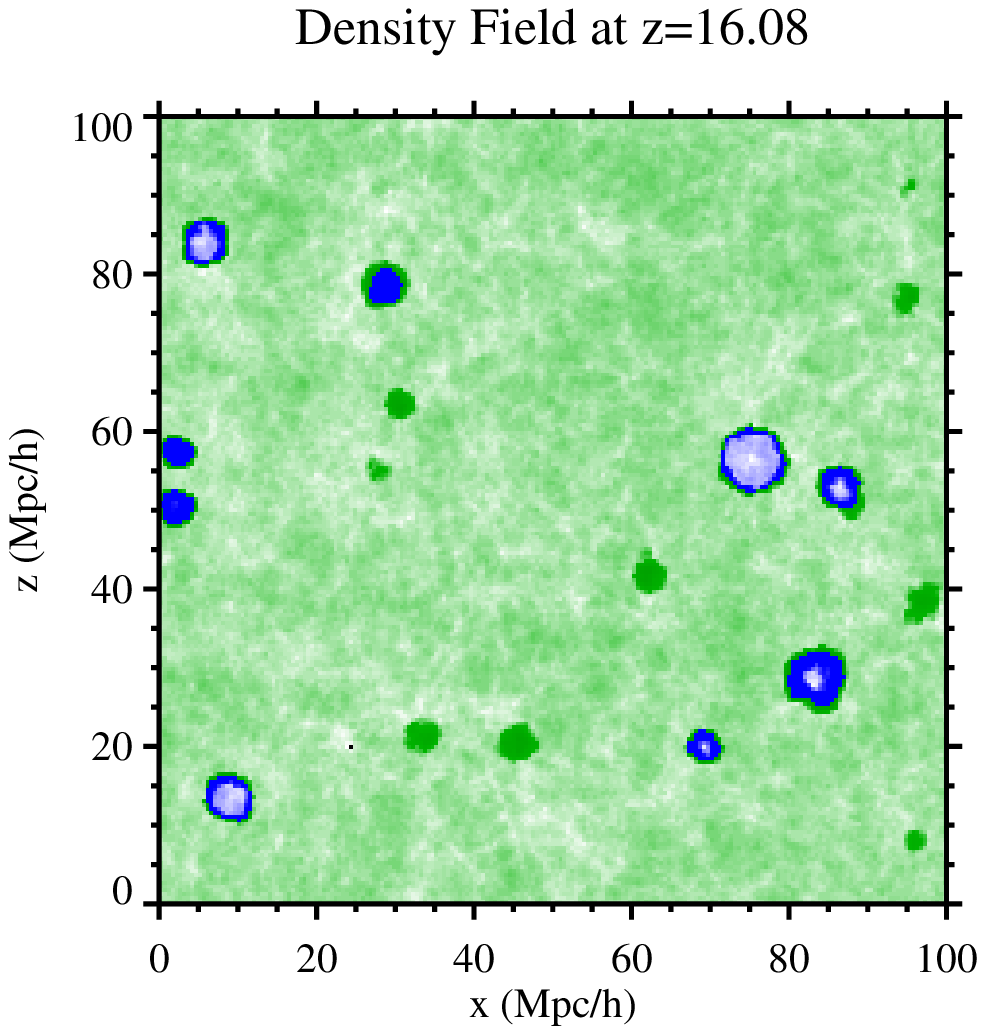}
\includegraphics[width=2in]{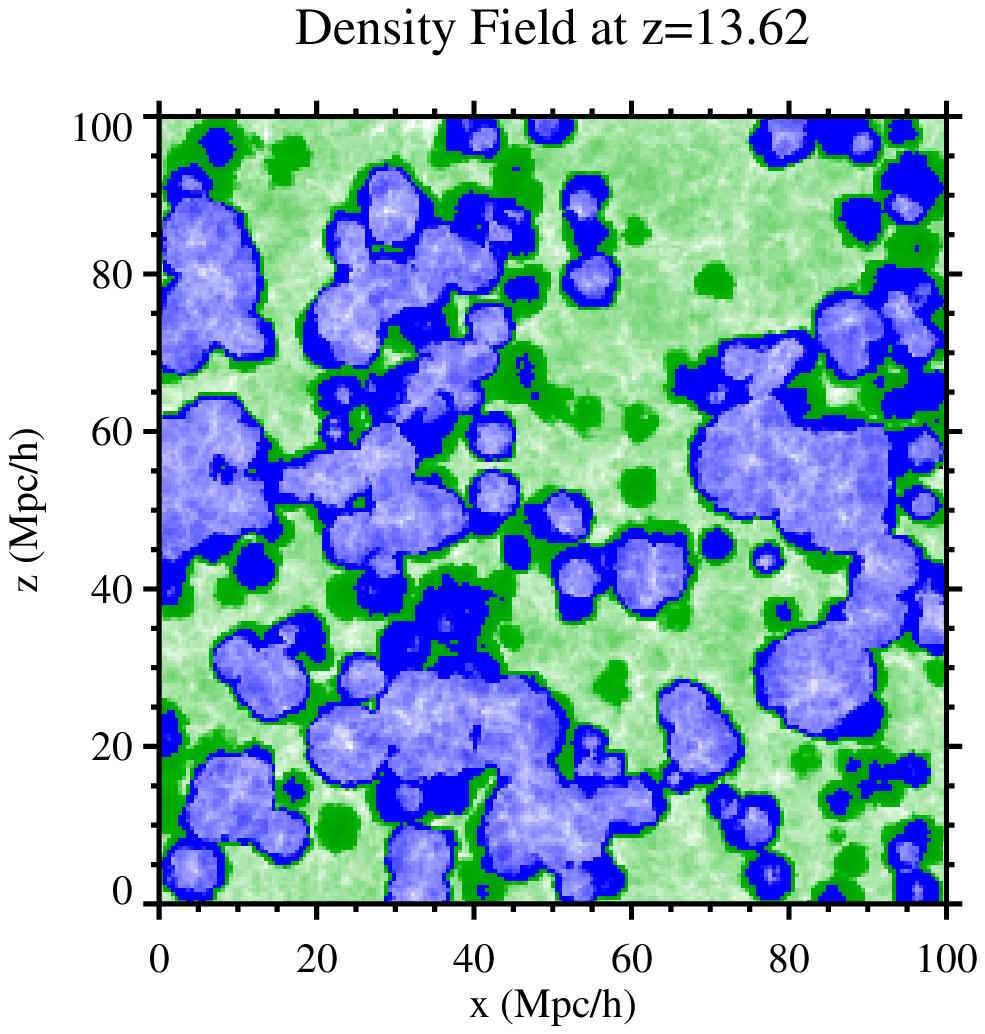}
\includegraphics[width=2in]{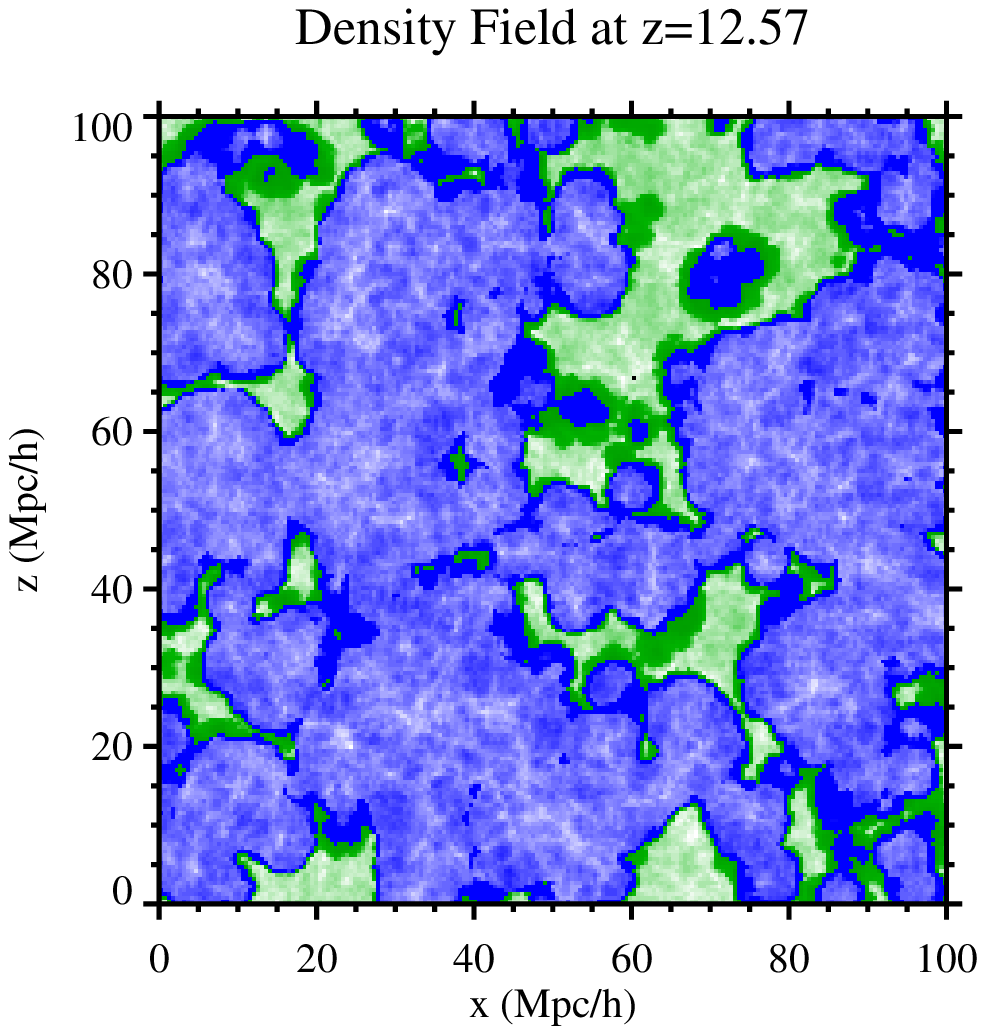}
\includegraphics[width=2in]{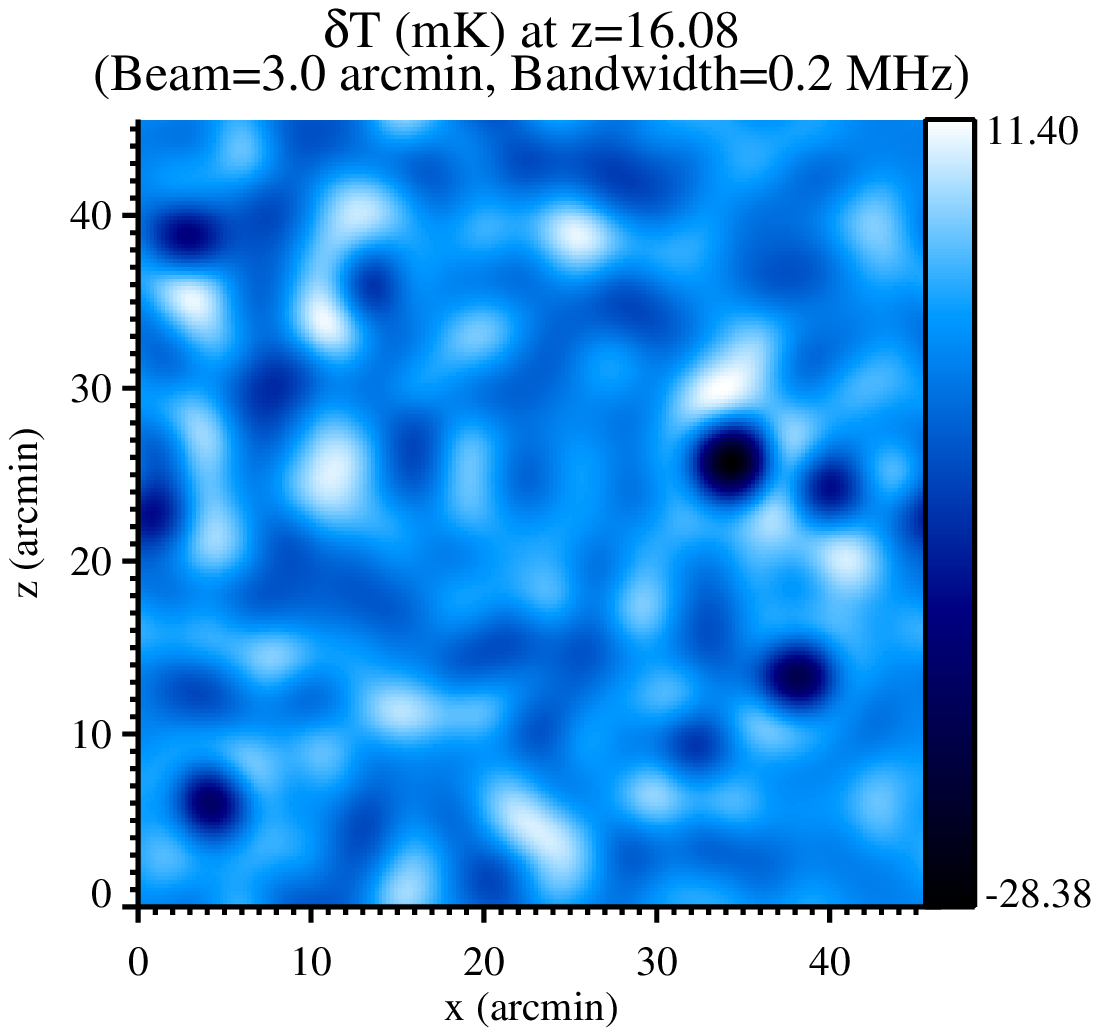}
\includegraphics[width=2in]{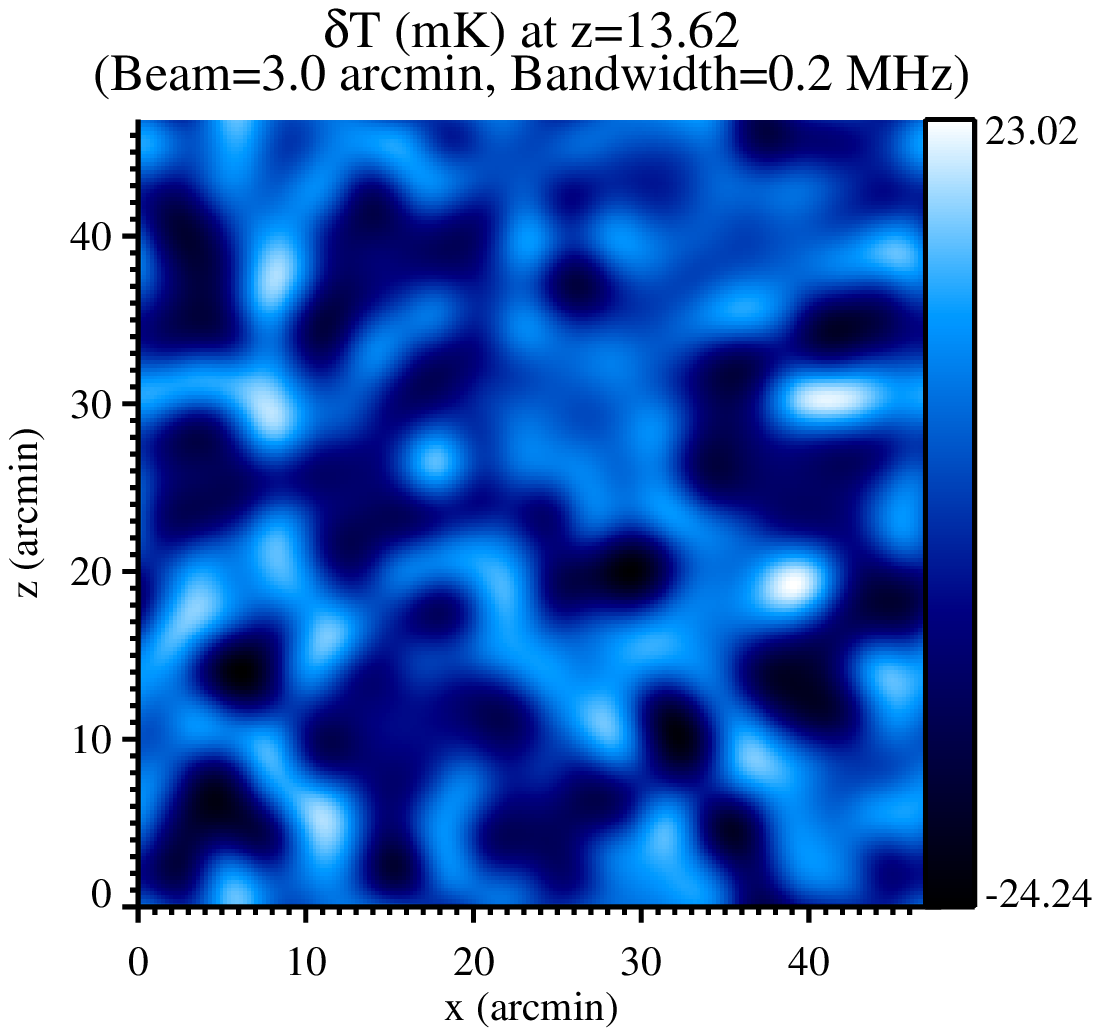}
\includegraphics[width=2in]{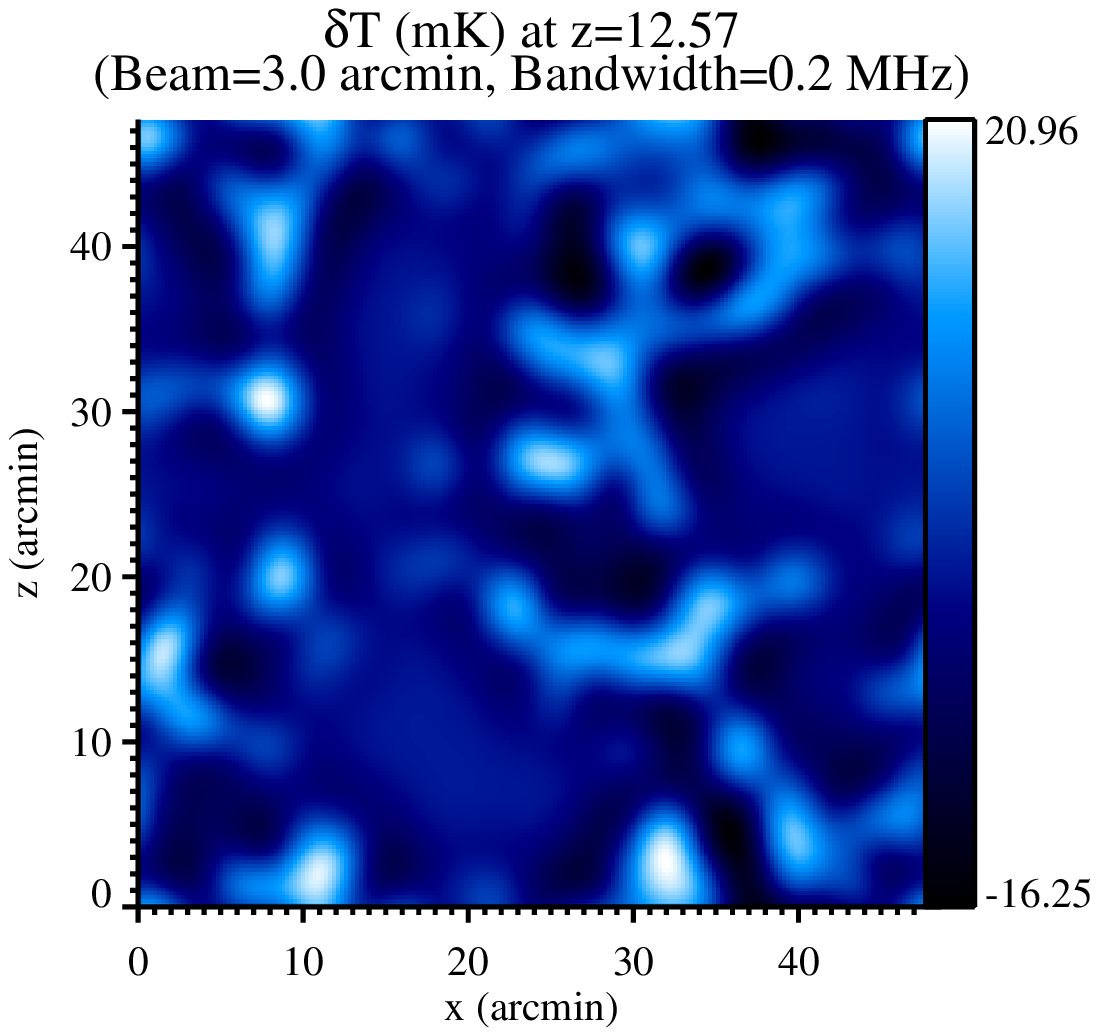}
\includegraphics[width=2in]{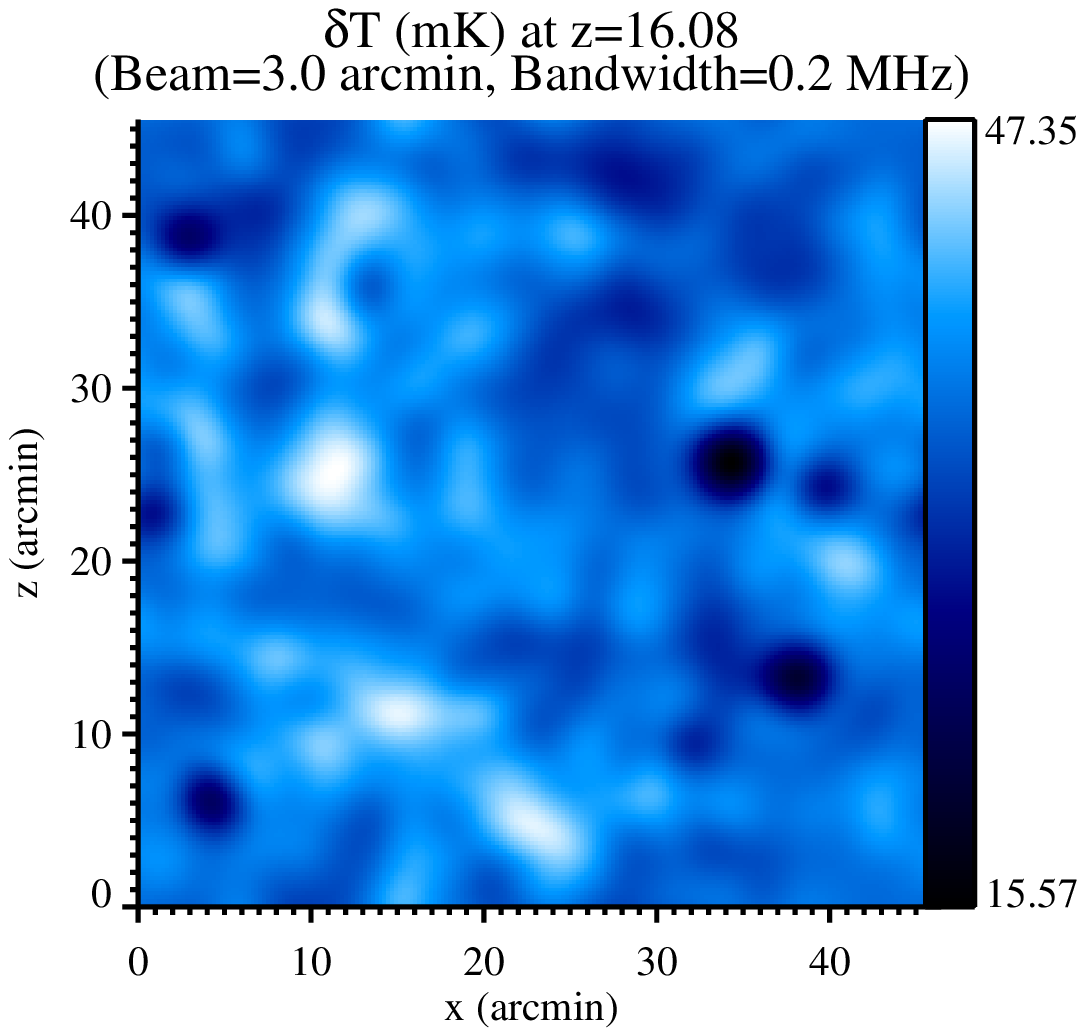}
\includegraphics[width=2in]{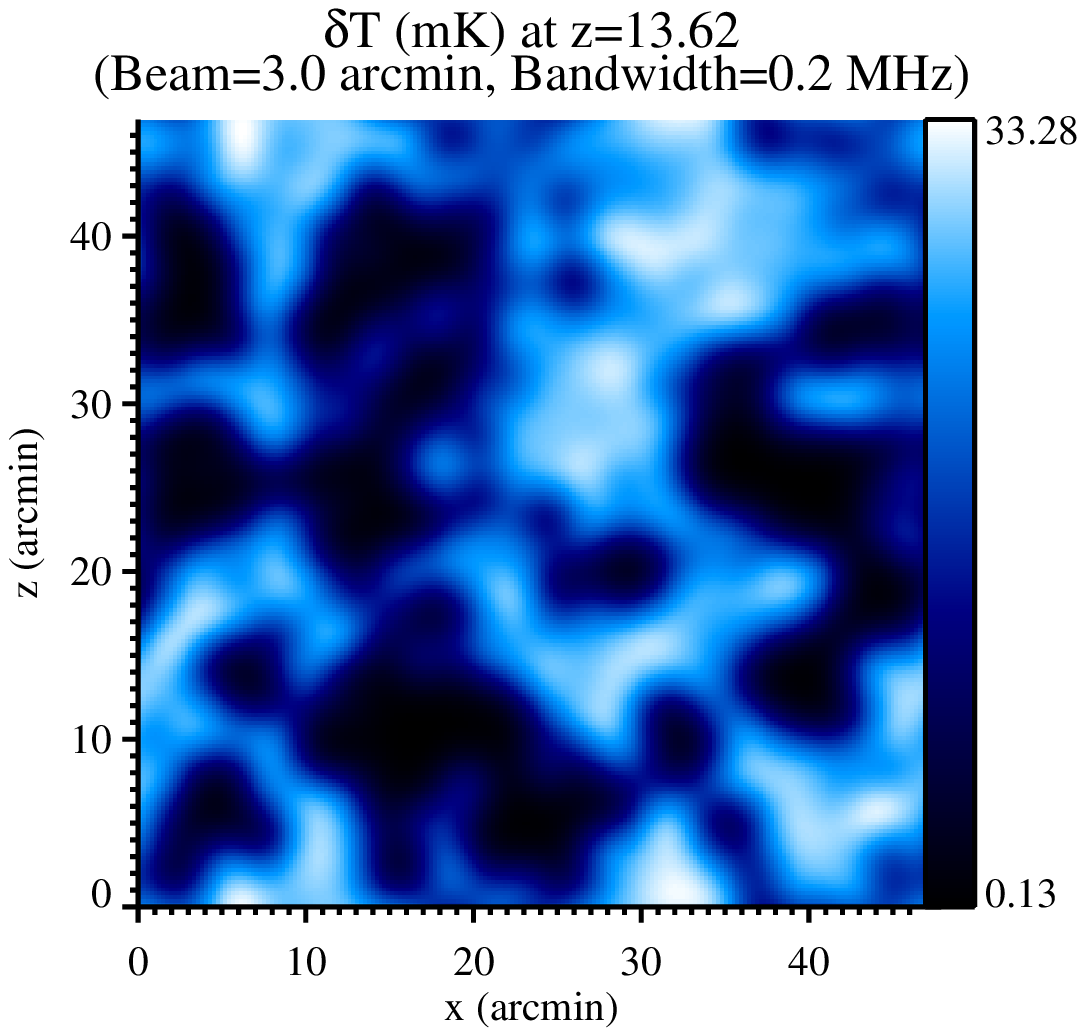}
\includegraphics[width=2in]{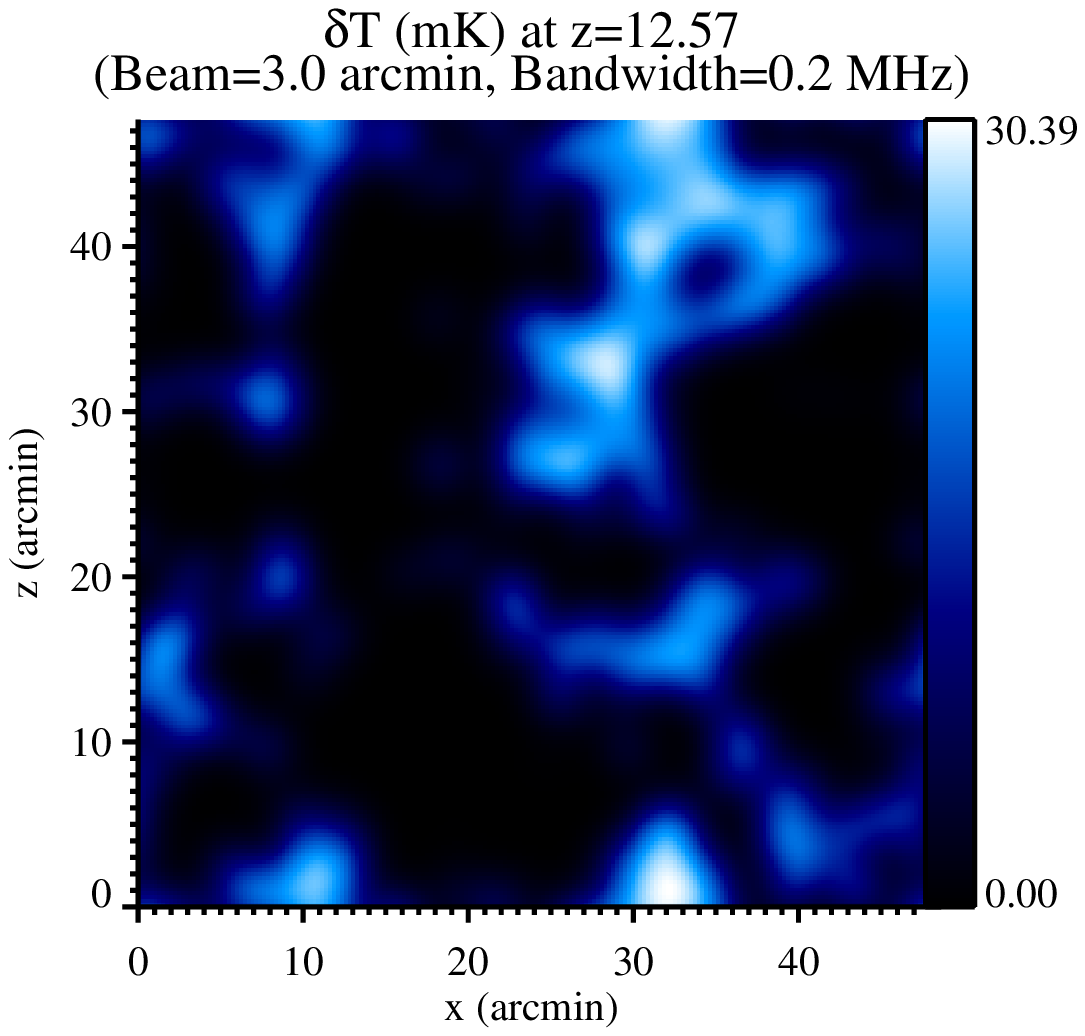}
\caption{The geometry of reionization seen at redshifted 21-cm line at
  redshifts $z=16.08, 13.62$ and 12.57 for f2000. Top row panels: a
  thick slice (corresponding to 0.2~MHz in frequency) showing H~II
  regions (blue) superimposed on the density field (green). {The
  ionized density increases from dark blue to white, the neutral
  density from dark green to white. The dark green and dark blue
  patches are those areas in the thick slice that contain a
  mixture of ionized and neutral material.} Middle row panels: The same
  reionization topology as seen at 21-cm with $3'$ FWHM compensated
  Gaussian beam and 0.2~MHz bandwidth.  Bottom row panels: The same as
  the middle panels, but with a Gaussian beam of the same FWHM.
\label{eor_topology_fig}}
\vspace{0.5cm}
\end{center}
\end{figure*}

In what follows we approximate the synthesized beam of angular size 
$\Delta\theta$ with a compensated Gaussian filter, given by
\begin{equation}
W_{\rm CG}(\theta)
=\frac{1}{2\pi\sigma^2}\left(1-\frac{\theta^2}{2\sigma^2}\right)
  \exp\left(1-\frac{\theta^2}{2\sigma^2}\right),
\end{equation}
with a FWHM equal to $\Delta\theta=2\sigma\sqrt{2(1-{\rm LambertW}(e/2))}$
(${\rm LambertW}(e/2)\approx 0.685$). The compensated Gaussian has a shape in
Fourier space whose real part, $k^2\exp(-k^2/2)$, approximates well the actual
observational beam shape of a compact interferometer (often referred to as
`dirty beam'), being insensitive to large scale features. For this reason 
we use the compensated Gaussian in most of our analysis. Its average value 
is zero, with a Gaussian-type peak in the middle, surrounded by a negative 
through. As a consequence, data convolved with this beam will have both 
positive and negative values. In some cases we also use two other beam 
shapes, a Gaussian with FWHM of $\Delta\theta$ and a top-hat function of 
width $\Delta\theta$. These two beam shapes are simple and widely used in 
the literature. We use them in a few cases below, and when we do so we
explicitly specify it. For reference we show plots of the angular power 
spectra of all three beam types in Fig.~\ref{beams_ps}. The frequency bandwidth
integration is always done with a tophat function, which is generally a good 
approximation to the actual integration used in observations.

\begin{figure*}
\begin{center}
\includegraphics[width=2.3in]{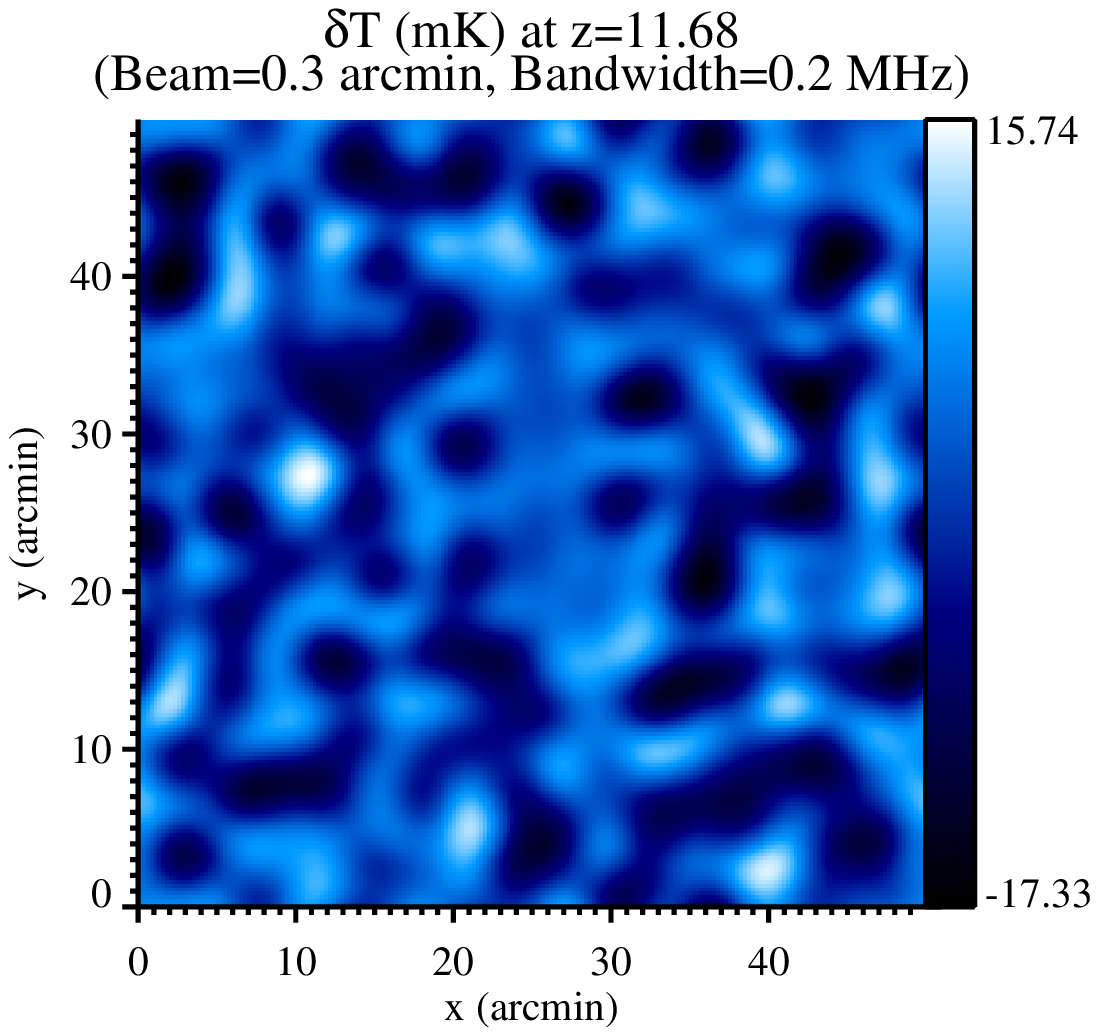}
\includegraphics[width=2.3in]{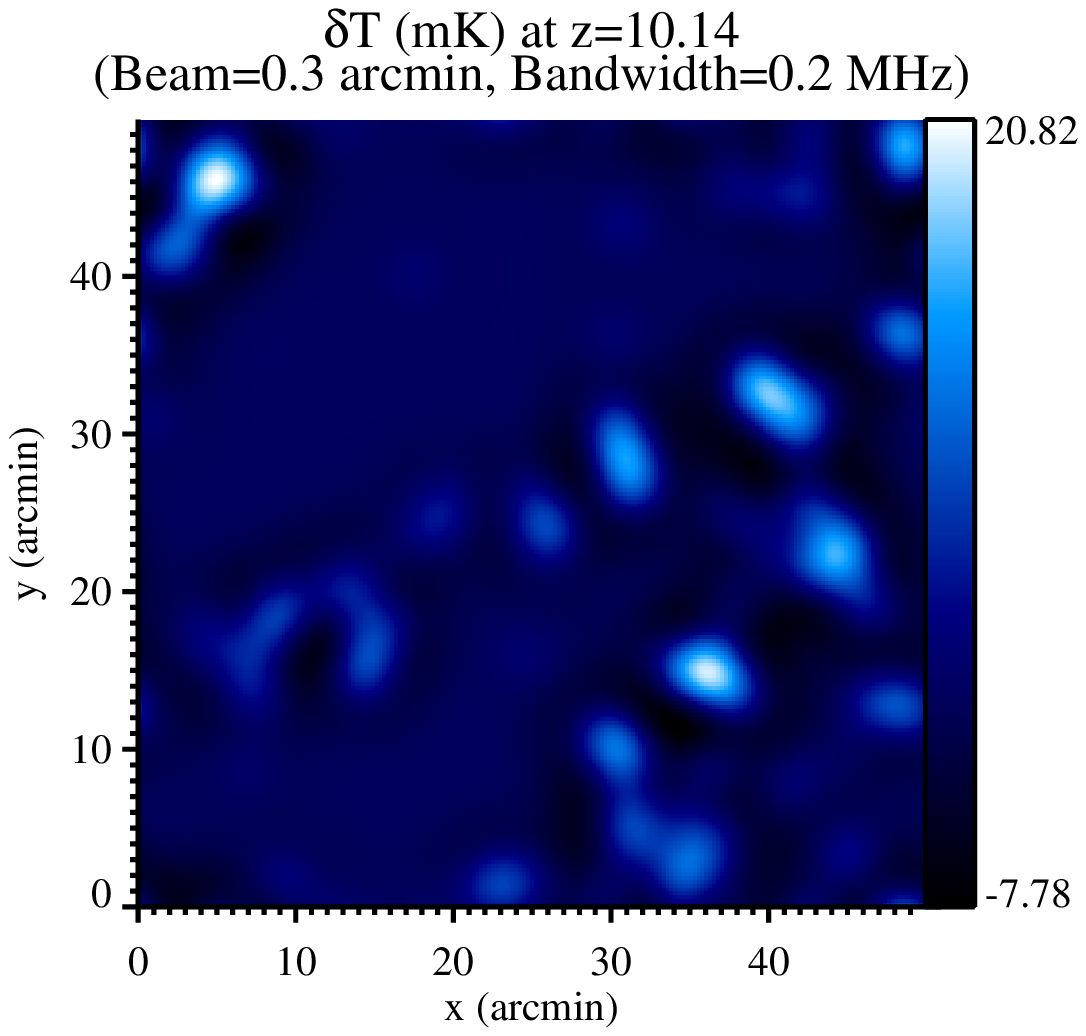}\\
\includegraphics[width=2.3in]{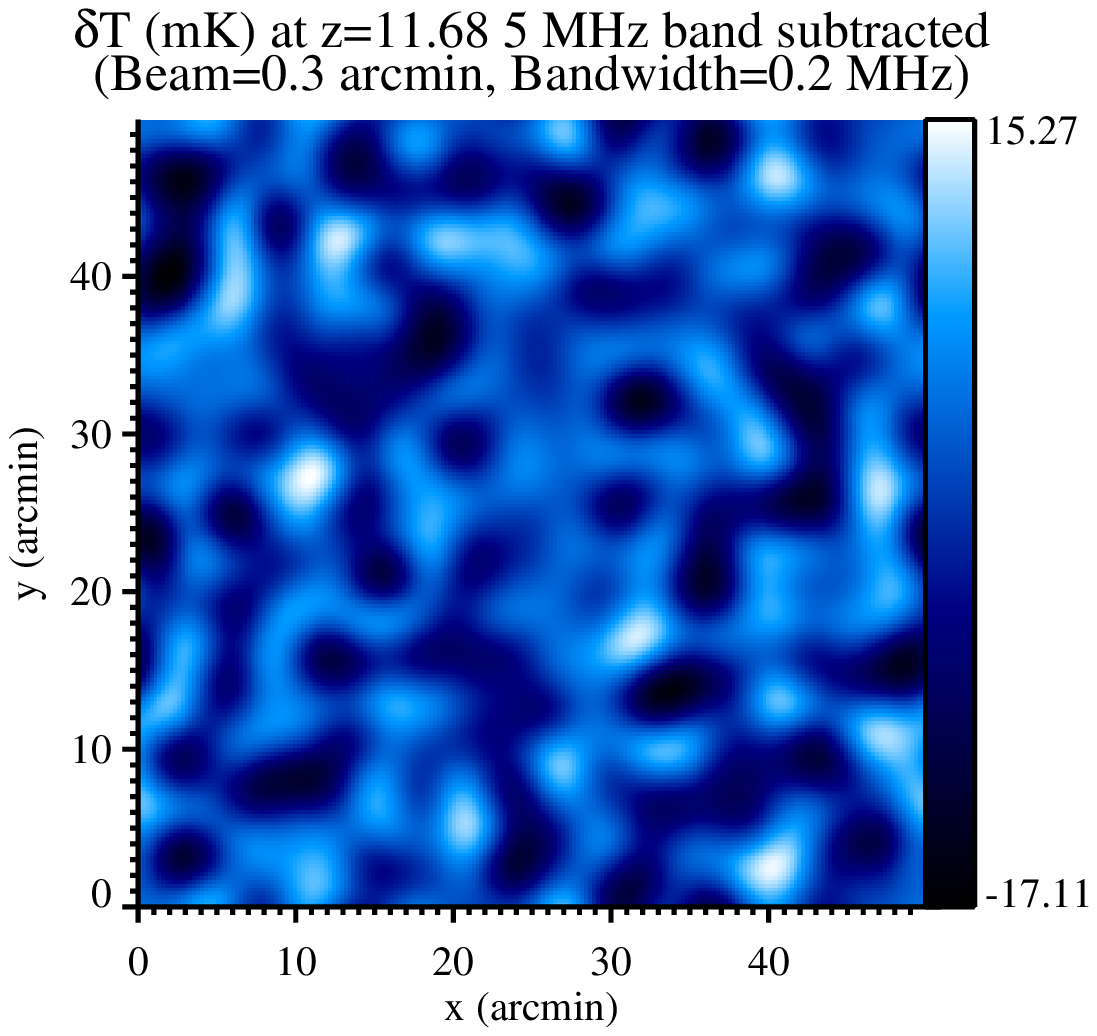}
\includegraphics[width=2.3in]{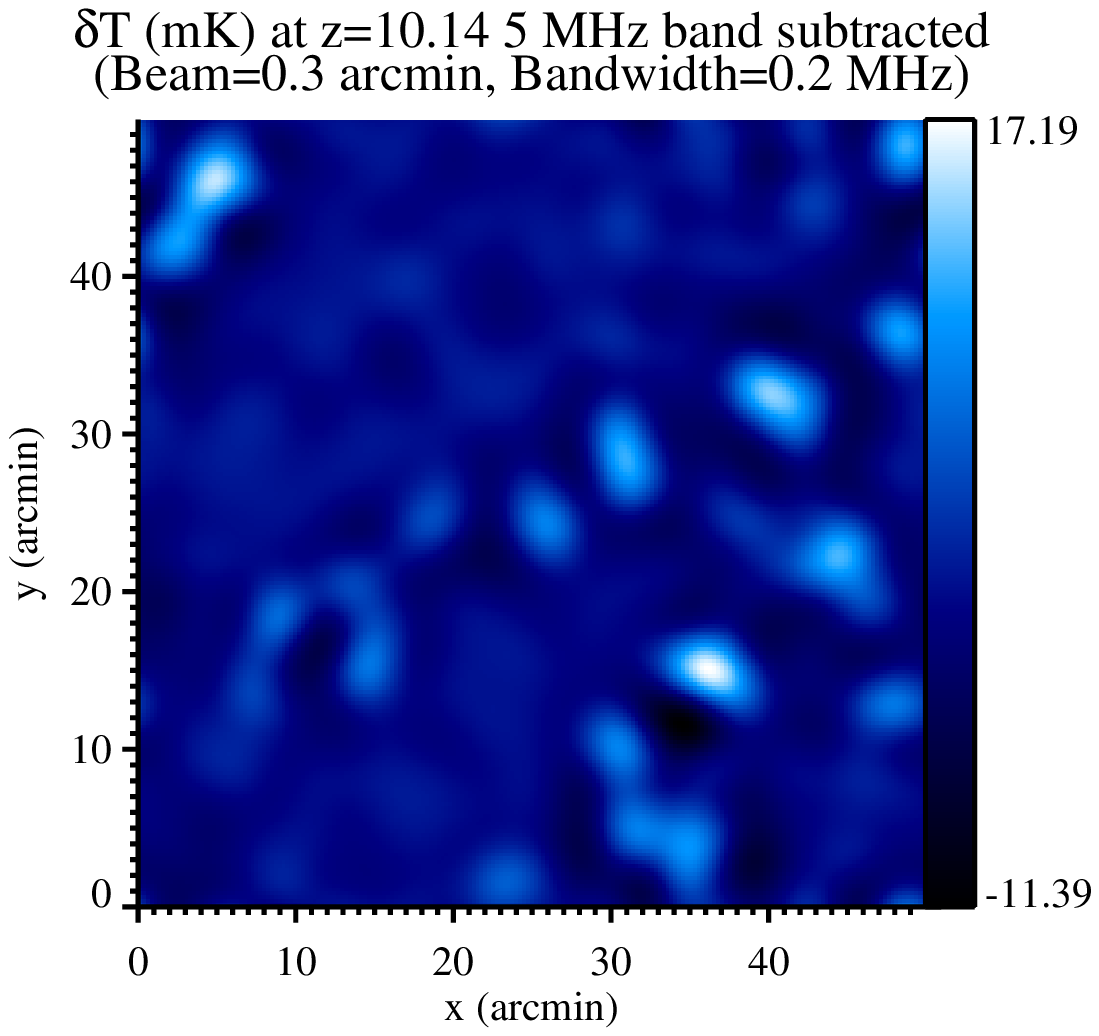}
\caption{(top) Sample 21-cm sky maps at redshifts $z=11.68$ and 10.14 from f250
  and (bottom) the same maps, but after subtracting a wide, 5~MHz band from it. 
  Subtracting the signal from such a wide band should remove most of the 
  foregrounds without affecting the underlying redshifted 21 cm signal.
\label{subtracted_maps_fig}}
\end{center}
\end{figure*}

In Fig.~\ref{eor_topology_fig} we show sample 'thick' slices of the
density fields (green/white) with the ionized regions shown in blue, at three
redshifts, taken from f2000 A thick slice contains the average of several slices corresponding
to a given band in frequency, to allow comparison to the images with 
finite bandwidth below. We show thick slices for redshifts $z=16.08$, early 
in the evolution (mass-weighted ionized fraction $x_m=0.026$), at $z=13.62$ 
when the volume is about half-ionized ($x_m=0.52$), and at at $z=12.57$, when 
most of the volume is already ionized, but significant neutral gas patches 
remain ($x_m=0.80$). Below each of those slices we show the 
corresponding 21-cm emission maps, each integrated over a bandwidth 
$\Delta\nu$=200~kHz and convolved with a compensated Gaussian beam of FWHM 
3\arcmin. The third row shows the images for a Gaussian beam of
the same FWHM. For simplicity, when integrating over the bandwidth we
neglected the redshift evolution and distortions over that bandwidth, which is justified for
such small $\Delta\nu$. As explained above, the compensated Gaussian beam
produces negative values since its average is zero.

An important point to note is that while the minimum and maximum values change
over time, the full range from the brightest to the least bright pixel in this
redshift interval is roughly independent of the redshift and remains around
30~mK for a Gaussian beam and around 40 mK for a compensated Gaussian.
However, this is not the case very early, when the reionization has barely
started and much later, close to overlap; in at those times the fluctuations
are less pronounced. This happens because during these epochs the fluctuations
are mostly dictated by the fluctuations of the underlying density field,
rather than by the patchiness of reionization, which strongly dominates the
fluctuations at the intermediate times. The images illustrate how well the
density fluctuations and the H~II regions would show on a realistic 21-cm sky
map. In spite of the significant level of smoothing, all large-scale features
remain clearly visible in the maps. The larger H~II regions are apparent as
holes in the 21-cm radiation, and the neutral density peaks correspond to
strong emission peaks. Some smaller, isolated ionized regions, with sizes
below the beam smoothing length disappear from the map, particularly for the
Gaussian smoothing. However, both the density fluctuations in the neutral
regions and the large-scale patchiness of reionization remain clearly visible.
The compensated Gaussian has much less power on large scales than the Gaussian
beam, but more power on smaller scales, and therefore shows the detailed
features better. The weaker 21-cm signatures of the mostly-ionized structures
inside the H~II regions are washed out and would require significantly higher
spatial resolution and sensitivity in order to be detected.

\begin{figure*}
\begin{center}
\includegraphics[width=2in]{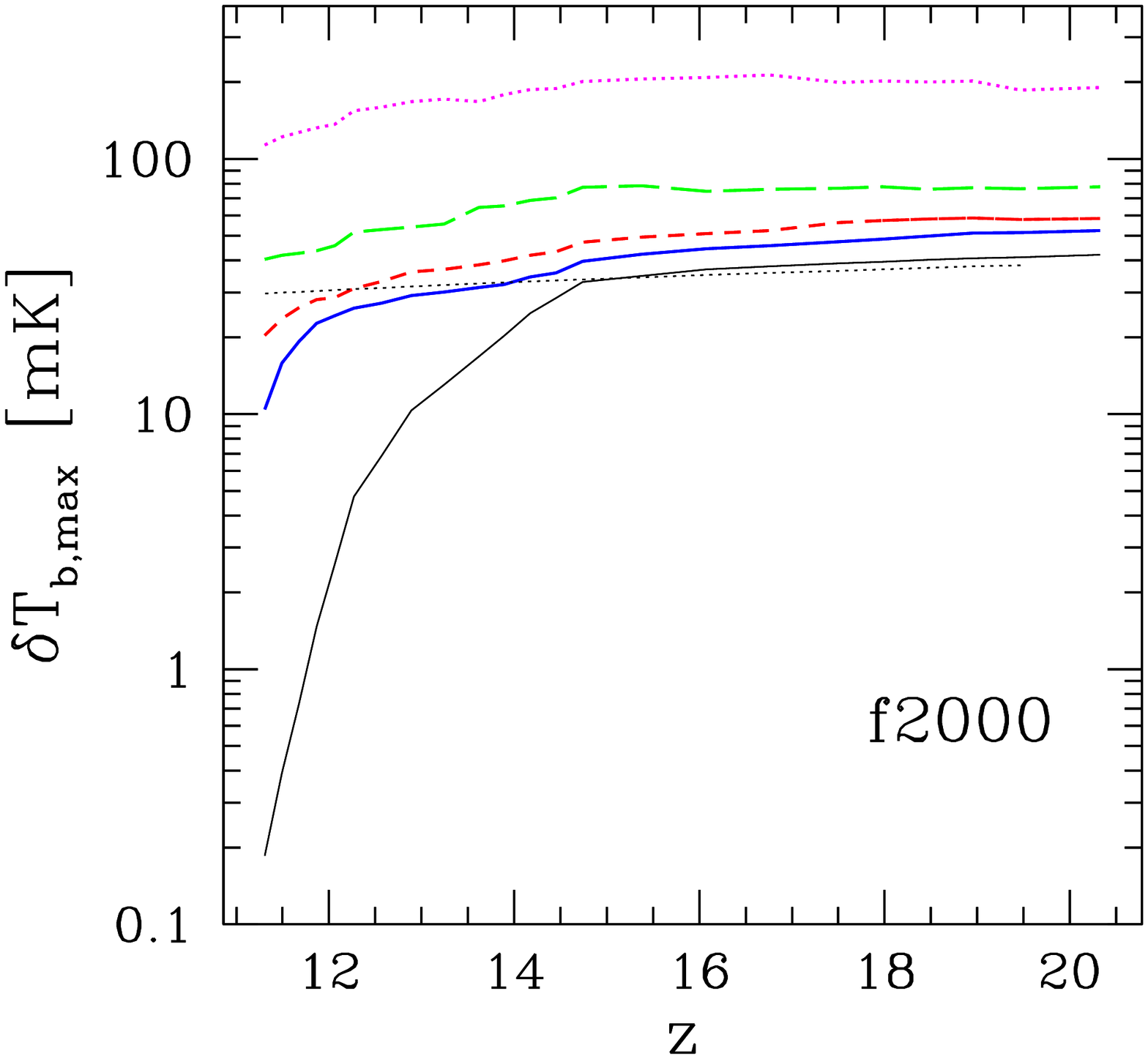}
\includegraphics[width=2in]{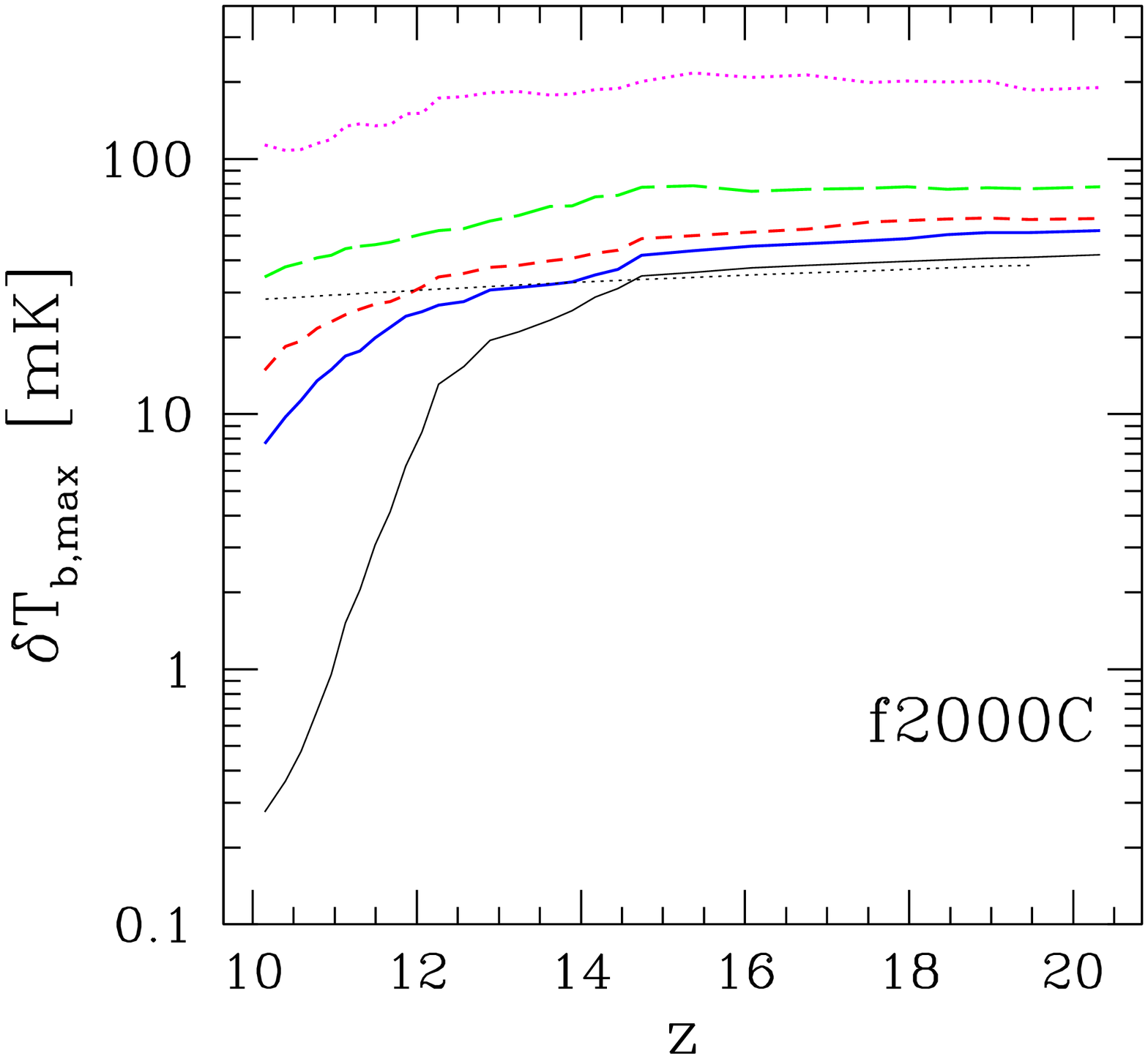}
\includegraphics[width=2in]{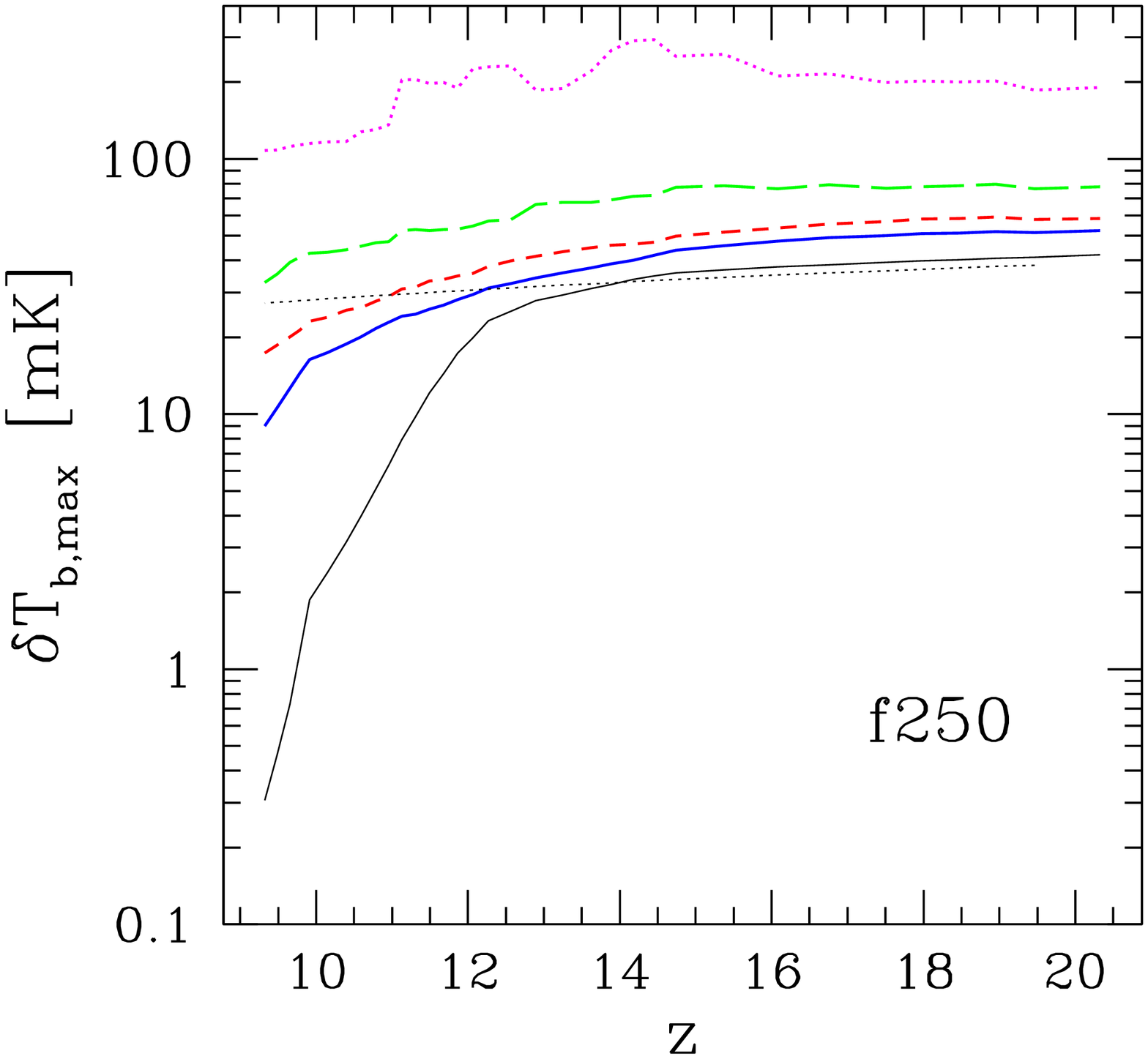}
\caption{The brightest peak in the simulation box as a function of redshift, 
  for (from left to right, simulations f2000, f2000C, f250, as labelled).  
  Shown are the maximum pixel value of the differential brightness 
  temperature, $\delta T_{b,max}$ vs. redshift $z$ for several beam-sizes and 
  bandwidths, as follows, $(\Delta\theta_{\rm beam},\Delta\nu_{bw})=(6,0.4)$ 
  (blue, solid), (3,0.2) (red, short-dashed), (1,0.1) (green, long-dashed) 
  and the full resolution of our simulation (i.e. corresponding to single 
  simulation cells; magenta, dotted), where the beam sizes are in arcminutes 
  and the bandwidths are in MHz. For reference we also show the mean 
  differential brightness temperature over the whole box (thin, black, solid) 
  and the same, but if the gas were fully-neutral (i.e. if no reionization 
  occurred; thin, black, dotted).
\label{maxima_fig}}
\end{center}
\end{figure*}

Although obtaining images of 21-cm emission would be the ideal
observational result it is not expected that such maps will be easily
extracted in the near term, due to the sensitivity limits of the
planned observations, and especially due to the strongly dominant
foregrounds. In order to eliminate the foregrounds effectively one has
to make use of their frequency characteristics. The issues of
foregrounds is a complicated one, and goes beyond the scope of this
paper. {However, if one naively approximates the foregrounds as a
slowly-varying power-law in frequency with smooth angular variations,
we can simply subtract the mean signal over a given band $\Delta\nu$.}
Here we take $\Delta\nu$ to be 5~MHz. Such an approach would average
out any small random variations of the spectral index, as well as
subtract the main, power-law-like component.

Figure~\ref{subtracted_maps_fig} shows two sample maps (using a
compensated Gaussian beam; top images) and the corresponding maps with
a 5~MHz band subtracted (bottom images). These correspond to the
half-ionized point ($z=11.68$, left panels) and the late phase
($z=10.14$, right panels) of our f250 simulation. The subtraction
process leaves the structures in the map largely intact, demonstrating
that such a subtraction procedure would not affect the signal
substantially, while at the same time eliminating much of the assumed
foregrounds. How well such a procedure would work in practice would of
course depend on how well-behaved the foregrounds are, {and in reality more 
intricate methods surely are needed.}

Obtaining detailed 21-cm sky maps would give us a very rich set of data on
both early structure formation and the progress of reionization. However,
extracting these from the noisy data dominated by strong foregrounds will not
be easy. Thus, in the following sections we first consider the detectability 
of individual large features and then statistical measures of the expected
signal, both of which should be significantly easier to obtain from the
observational data, and so should be the first goal of the observations.

\section{Individual features}
\label{indiv_sect}

\subsection{Rare, bright peaks}
\label{maxima}
The brightest emission peaks for a given beam and bandwidth are possibly 
the 21-cm emission features easiest to detect. In Figure~\ref{maxima_fig} we
show the maximum pixel value of the differential brightness temperature,
$\Delta T_{b,max}$ versus redshift $z$ for beam-sizes and bandwidths
$(\Delta\theta_{\rm beam},\Delta\nu_{bw})=(6,0.4)$ (roughly corresponding to
the ones relevant to PAST), (3,0.2) (``LOFAR''), (1,0.1) (``SKA'') and
the full resolution of our simulation (corresponding to single cells),
where the beam sizes are in arcminutes and the bandwidths are in MHz. In order
to compare to the mean signal and the signal without reionization, we 
use a Gaussian beam here, since the compensated Gaussian has a zero mean by
definition.

The first thing to note is that the values and the shape of the curves do not
depend significantly on the reionization history, except for the a shift in 
redshift, by about $\Delta z=1.5$ (2) between f2000 and f2000C (f2000 and
f250). The magnitudes of the brightest peaks is fairly high, at several tens 
of mK even in the beam- and bandwidth-smoothed cases, and well over 100 mK 
for full resolution. The magnitudes generally drop with redshift, but only 
mildly and the curves stay largely flat over the whole redshift range. The 
higher resolution and sensitivity of SKA would provide a significant 
improvement, while the difference in the signal between the LOFAR and PAST 
scales is rather marginal.

These brightest peaks generally correspond to high-density, still
neutral regions, i.e.\ regions that are on the verge of forming galaxies, 
who in turn would start ionizing their surroundings. Hence, while the 
magnitude of the peaks do not vary much with redshift, the location in the box 
at which it is found does vary significantly and the brightest peaks at one 
redshift become deep troughs at later times. For reference we also show the 
mean simulated 21-cm emission signal, as well as the mean signal if we assume 
the volume to be completely neutral (i.e.\ if reionization never happened). 
The mean signal from the box is well below the maximum one at all redshifts, 
even for maximum smoothing, and it decreases much more steeply with redshift, 
thus the difference from the peaks grows significantly with time, from factors 
of 1.2-6 (depending on the smoothing) at $z\sim16$, up to 35-320 at late times
($z\sim11.5$). Notably, the peak signals are also generally higher than the
mean signal for a neutral volume.
This indicates that the peak signals always 
come from overdense regions, even though, as we showed in Paper I, the 
reionization is clearly inside-out, i.e. the high-density peaks get ionized 
preferentially compared to the voids. For high levels of smoothing (i.e. 
large beam sizes and bandwidths) the peak signal drops below the corresponding 
fully-neutral gas signal since by that time our volume is more than
half-ionized and hence any beam that samples much of the box would inevitably 
include significant ionized regions. This is a consequence of the
high bias of the density peaks (and sources) at high redshift, which means that
the maximum-signal peak inevitably is close to other density peaks that have
been ionized already.

\begin{figure*}
  \begin{center}
    \includegraphics[width=\textwidth]{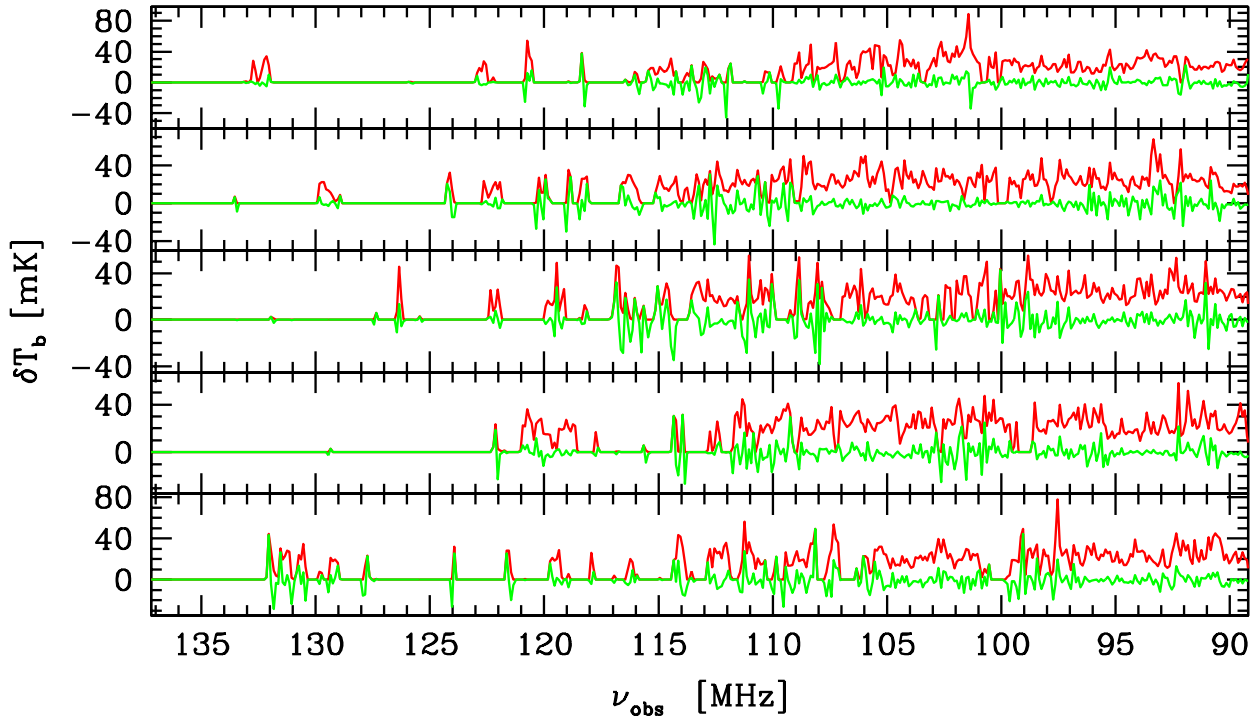}
    \caption{Sample line-of-sight 21-cm spectra obtained from our
      simulation data (simulation f250). The figure shows the redshift
      distorted spectra (red) and the difference between the distorted
      and the undistorted spectra (green).
      \label{1D_spectra_fig}}
  \end{center}
\end{figure*}

\begin{figure*}
  \begin{center}
    \includegraphics[width=\textwidth]{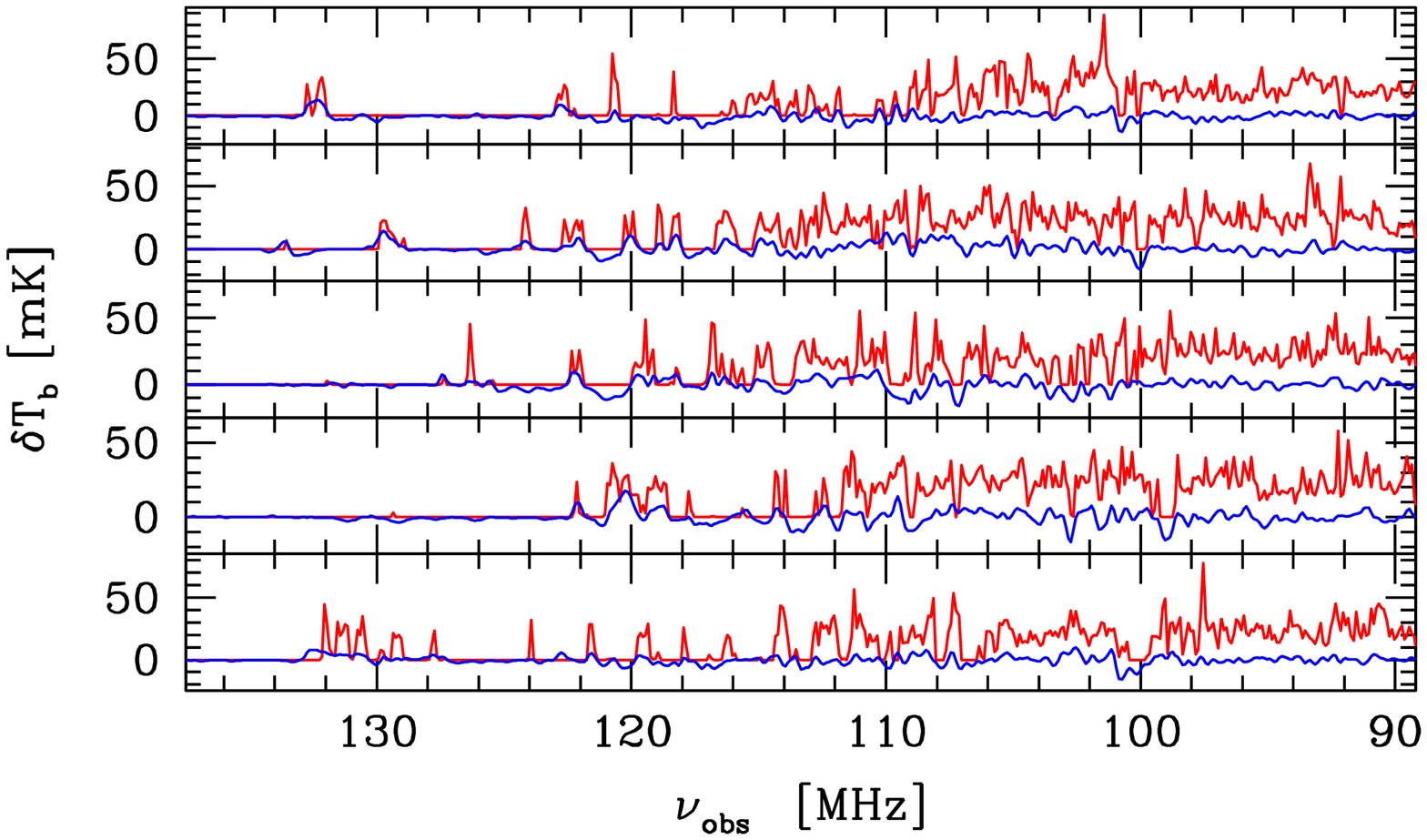}
    \caption{Sample line-of-sight 21-cm spectra obtained from our
      simulation data (simulation f250). Shown are the full-resolution
      (red) and the beam- and frequency-smoothed spectra (blue). For
      the latter we used a compensated Gaussian beam with a FWHM of
      3$^\prime$ and a bandwidth of $0.2$~MHz.
      \label{1D_spectra_fig_smooth}}
  \end{center}
\end{figure*}

Searching for the relatively rare, bright peaks could thus be an efficient way
to detect the signatures of reionization, even in its late stages when the
universe is already more than 99\% ionized. This would make these stages
observable to radio arrays which are strongly affected by interference in the
FM-band, such as LOFAR and the GMRT, since they correspond to frequencies well
above the FM band. The statistics of these rare events is discussed in
Sect.~\ref{nongauss_sect}.

\begin{figure}
  \begin{center}
    \includegraphics[width=3.5in]{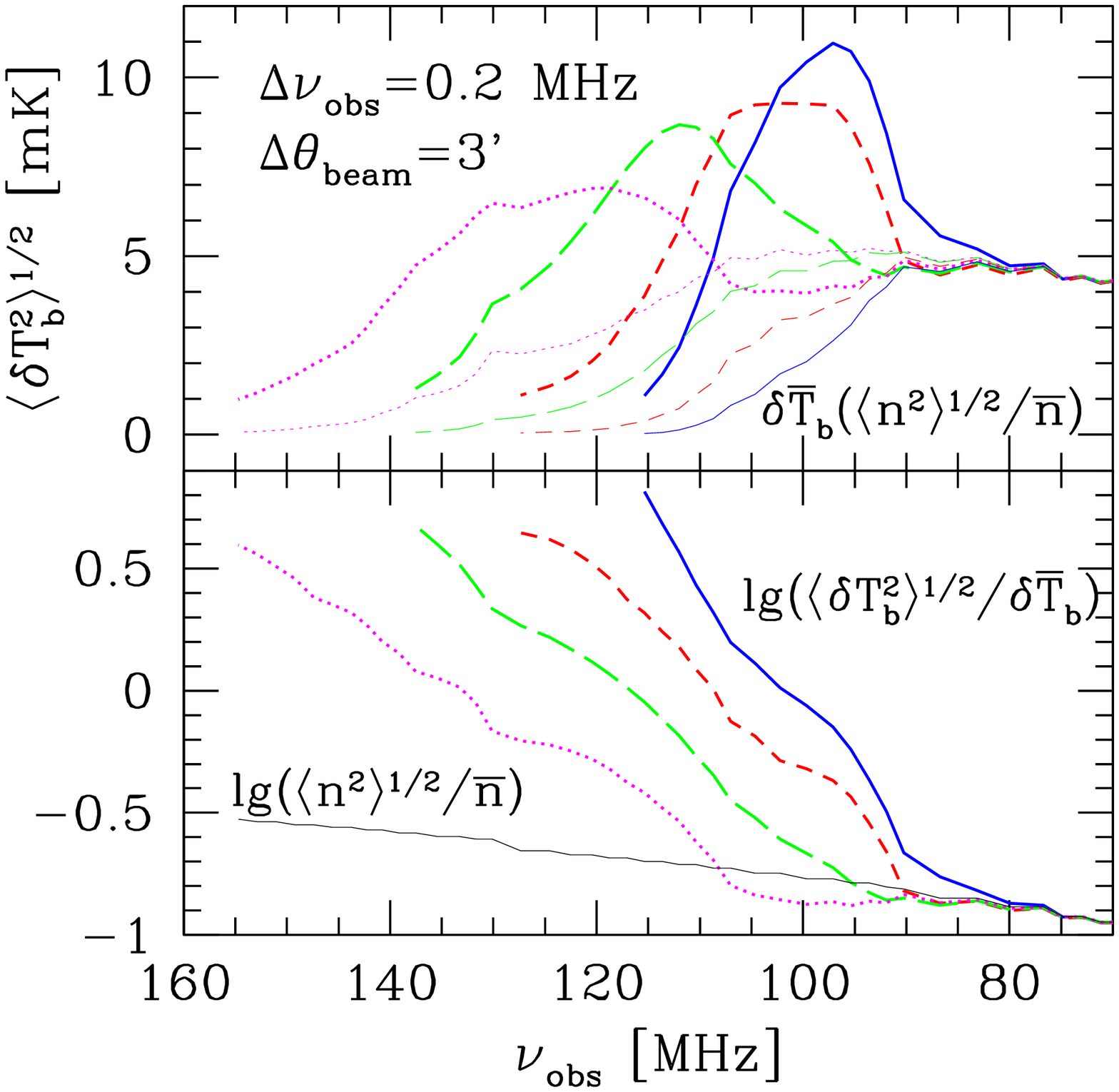}
    \caption{(top) Rms fluctuations of the differential brightness
      temperature, $\langle \delta T_b^2\rangle$ vs. observed
      frequency, $\nu_{\rm obs}$ for f2000 (solid, blue), f2000C
      (short-dashed, red), f250 (long-dashed, green) and f250C
      (dotted, magenta). We also show the corresponding $\delta
      \bar{T}_b(\langle n^2\rangle^{1/2}/\bar{n})$, i.e. the
      temperature fluctuations if they were following the fluctuations
      of the density, normalized to their respective means (thin
      lines, same line types and colours for the cases). (bottom) the
      density (thin, solid, black) and differential brightness
      temperature fluctuations relative to their respective means
      (thick lines, same line types and colours for f2000, f2000C,
      f250 and f250C as in the top panel) vs. frequency.
      \label{fluctfig}}
  \end{center}
\end{figure}

\subsection{Line-of-sight spectra}
\label{LOSspectra}
Next we present line-of-sight (LOS) 21-cm emission spectra obtained from our
simulations. These can in principle probe in detail the neutral structures at
high-redshifts along that LOS, as well as detect the H~II regions as deep
troughs in the spectra. In order to obtain spectra over large frequency bands
we use the technique discussed in Sect.~\ref{global_evol_sect} for deriving the
evolution over the complete redshift and time intervals of our simulations.
We present the full-resolution spectra (essentially just LOS cuts through the
data shown in Fig.~\ref{pencil}), as well as the corresponding beam- and
bandwidth-smoothed spectra for our simulation f250, since its overlap is
latest, thus it extends furthest into the most easily observable redshift 
range.

The full-resolution (at resolution $\sim30$ kHz, corresponding to one
computational cell) spectra are shown in Figure~\ref{1D_spectra_fig}.
The red line shows the spectra including redshift distortions due to the bulk
peculiar velocities, and the green line is the difference between the 
distorted and undistorted spectra. The redshift distortions introduce 
substantial changes in the spectrum, moving peaks to different frequencies 
(showing up in the difference as a deep dip next to a high peak), merging
the emission of some gas parcels, and spreading others out over a larger range
in frequency. The comparison shows that these are noticeable effects that 
should always be taken into account. These redshift distortions can in
principle be used to derive information about the underlying cosmology as well
as about reionization \citep{2005ApJ...624L..65B,2005astro.ph.12263M},
but this falls outside the scope of the current paper.

As was also noted above, at high resolution there are some very bright pixels,
with $\delta T_b>50$~mK, and these are fairly common during most of the
evolution, and generally at least one is found on every LOS. However, they
become increasingly rare after $z\sim12$ ($\nu_{\rm obs}\sim110$~MHz), since
by that time most of the high-density peaks have formed ionizing
sources. Conversely, at about that time the H~II regions become common and
start to be seen along most LOS as deep troughs. The size of these troughs
initially corresponds to the typical size of the ionized regions, $\sim10$~Mpc
comoving, or about 0.5~MHz. As discussed in detail in Paper~I around the time
the volume is half-ionized these ionized regions quickly start merging with
each other and start forming much larger ones, seen here are wider troughs, of
several MHz each. Most of the signal eventually disappears as the volume
becomes ionized.  However, as we already pointed out, this occurs at quite
different frequencies along different LOS. Along some LOS there are
essentially no neutral structures for $\nu>120$~MHz, while along others
significant neutral regions persist until $\nu\sim135$~MHz or later.

These fairly sharply-defined H~II regions suggest a local (on scales tens of
Mpc) equivalent of the ``Global step'' discussed in Sect.~\ref{mean_bg_sect}. 
While the last proved rather gradual, on a more local level the steps are be 
quite steep, from 30-50 mK down to almost 0 and back within a fraction of a 
MHz. 

In Figure~\ref{1D_spectra_fig_smooth} we show the effect of beam- and
bandwidth-smoothing on the spectra. The beam we use is again the compensated
Gaussian, which produces negative differential brightness values at the
minima. Clearly the smoothing reduces the signal considerably when most of
the material is still neutral ($\nu\lesssim100$~MHz), and the fluctuations
are entirely due to density fluctuations, at relatively small scales. Once
the first H~II regions begin to expand, they produce much stronger
fluctuations. These show up in the spectra, although the smallest ones are
still averaged out by the smoothing. The valleys corresponding to the H~II
regions are generally deeper than the neutral density peaks, so they should
be easier to detect. At later stages the isolated peaks are suppressed in
some cases, but often remain clearly visible, depending on their
characteristic size and environment. For these LOS and smoothing the signal
never varies more than a few tens of mK, but the main features of the
unsmoothed data are preserved and should still be detectable.

\section{Statistical signals}
\label{stat_sect}

\subsection{Spatial rms fluctuations}
\label{rms_sect}
The prominent individual 21-cm features we discussed in the previous section,
either rare, bright emission peaks or large individual H~II regions would
quite possibly be the first signatures of the reionization epoch to be
seen. Once such a first detection is made the next aim would be to put more
robust constraints on the progress and duration of reionization than the ones
currently available. At the next level of complexity, the data may be used to
derive the statistical measures of the fluctuations in the 21-cm signal. This 
could be achieved by considering a number of observational fields, thus 
suppressing the noise and systematics in the foregrounds. 

The fluctuations of the 21-cm EOR signal can come from several sources. Here
we concentrate on the dominant effect, the large-scale patchiness of
reionization. The fluctuations can be derived as a single number (rms), for
any given scale (defined by the observational beam and bandwidth), or in more
detail in the form of angular power spectra. We consider both here, starting
with the rms fluctuations. The rms fluctuations are the equivalent of the
global step discussed in Sect.~\ref{mean_bg_sect}, but observed with an
interferometer (which is insensitive to a mean global signal).

In Fig.~\ref{fluctfig} we show how the fluctuations in the mean signal
change with redshift/frequency (top panel) for f2000, f2000C, f250 and
f250C. These were calculated using a top hat beam of 3\arcmin\ and a
bandwidth of 0.2~MHz. The top hat was used here for computational
convenience, using a different beam would change the numbers slightly,
but would not affect the basic trends which we want to illustrate
here. Early-on the 21-cm fluctuations closely follow the gas density
fluctuations, since very little of the gas is ionized. Once some more
substantial ionized patches start appearing, with sizes roughly of
order of the beam- and bandwidth smoothing, the differential
brightness temperature fluctuations quickly raise and reach a maximum
of $\langle \delta T_b^2\rangle^{1/2}\sim10\,\rm mK$, after which they
drop as more and more of the universe becomes ionized, reaching
$\sim1$~mK at overlap. This behaviour is quite generic and occurs in
all cases presented here. However, the frequency, $\nu_{\rm max}$ (and
redshift) at which the maximum of the fluctuations is reached varies
significantly among the runs, from $\nu_{\rm max}\sim97$~MHz for f2000
up to $\nu_{\rm max}\sim120$~MHz for f250C.  The peak value of the
fluctuations decreases modestly for more extended reionization scenarios, from $11$~mK for
f2000, down to $7$~mK for f250C. {This is due to the fact that both
clumping and lower photon efficiency move power to smaller scales, below
the smoothing scale used here.} In all three cases the peak
fluctuations occur at the time when the universe is $\sim50\%$ ionized
by mass. The peaks are roughly symmetric in all cases and
significantly wider for the late-overlap scenario (f250) than for the
early-overlap one (f2000). {The increasing clumping $C(z)$ with time, 
combined with an increasing number of sources leads to a broader peaks for the 
models with redshift dependent clumping (f2000C and f250C).}

As a reference we also show the $\delta \bar{T}_b(\langle
n^2\rangle^{1/2}/\bar{n})$, i.e. the fluctuations for reionization scenarios
with the same $\delta\bar{T}_b(z)$ as the actual simulations, but assuming
that the process occurred homogeneously rather than patchily. In this last
case the fluctuations of the emission would have been solely due to the
fluctuations in the underlying density field. Clearly the patchy reionization
creates substantial extra power over a wide range of frequencies. The
difference is especially large close to overlap, where the mean signal
essentially disappears, while the actual patchy signal is $\sim1$~mK even at
overlap, when the mass neutral fraction is less than 1 per cent. The bottom
panel underlines this point further by showing the rms fluctuations of the
differential brightness temperature relative to its mean, $\langle \delta
T_b^2\rangle^{1/2}/\delta \bar{T}_b$, for all four cases, and the same for
the density fluctuations, $\langle n^2\rangle^{1/2}/\bar{n}$ (where the last
is the same for all cases, since they use the same underlying density
field). We see that the relative differential brightness temperature
fluctuations are more than an order of magnitude larger than the relative
fluctuations of the underlying gas density field.

\begin{figure*}
\begin{center}
\includegraphics[width=3.2in]{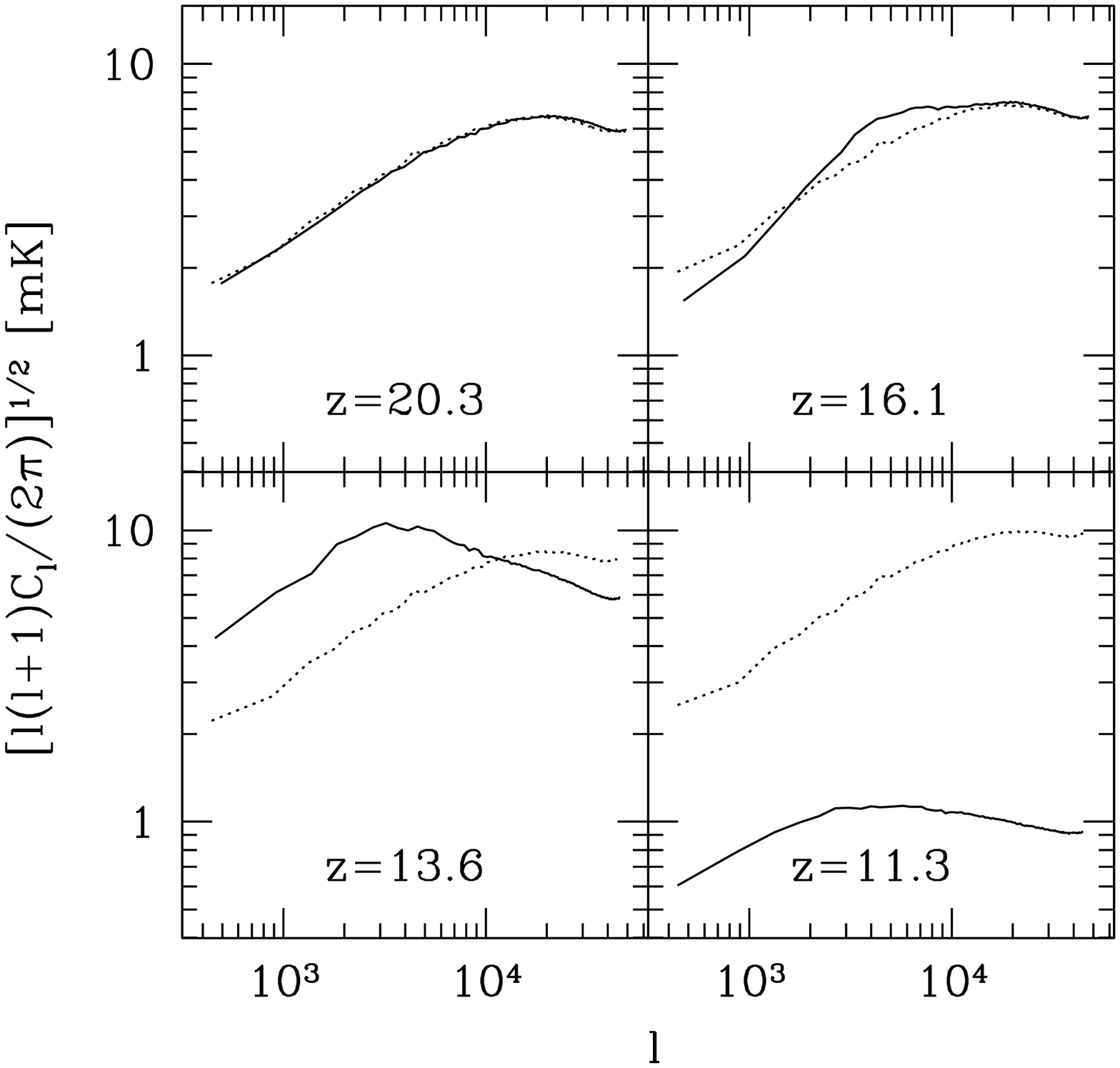}
\includegraphics[width=3.2in]{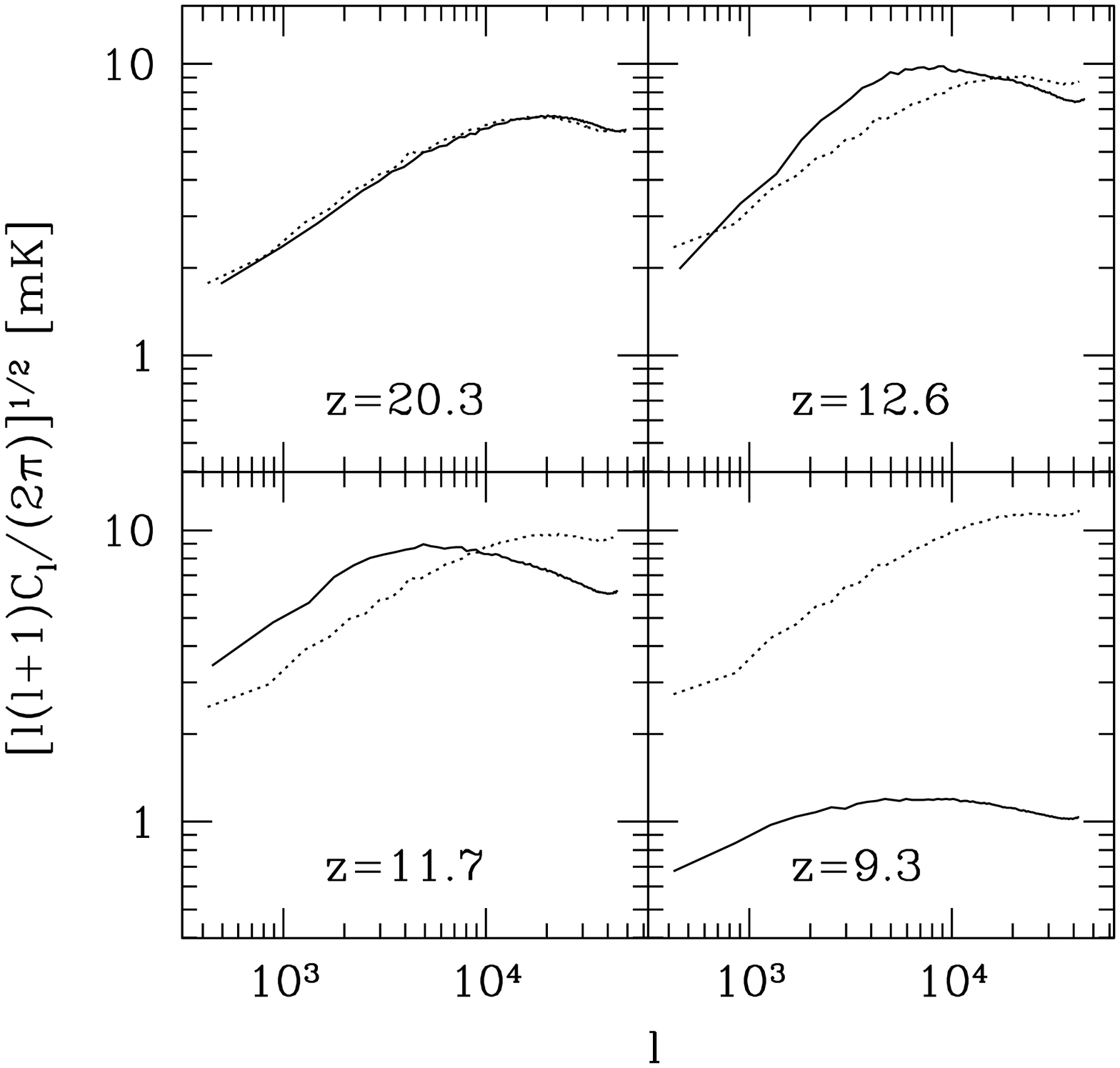}
\caption{2D angular power spectra of the differential brightness temperature
  fluctuations (solid lines) for f2000 (left) and f250 (right). For reference 
  we also show the corresponding fluctuations if there were no reionization
  (i.e. due just to the fluctuations of the underlying gas density field
  assuming all the gas were neutral).   
\label{pow_2d_fig}}
\end{center}
\end{figure*}

{Although less of a requirement than in the case of the global step, a
sudden decrease in the rms fluctuations should be easier to detect
observationally, since other, possibly confusing sources of fluctuations,
e.g.\ due to the instrumental setup or the foregrounds, are expected to
change only slowly with frequency}. Our f2000 simulation does indeed shows a
fairly steep drop over 15~MHz after the peak (but located in the FM band),
whereas the other cases display a more gradual change over 20--30~MHz.

\subsection{Power spectra}

The quantity closest to the actual radio array observational data is the
2D angular power spectrum. It shows how the power of the fluctuations is
distributed over the different angular scales.
We follow the convention used in CMB analysis and construct the angular
power spectrum expanded in spherical harmonics,
$[\ell(\ell+1)C_l/2\pi]^{1/2}$, where $\ell=2\pi/\theta$, with the angle
$\theta$ in radians. The results are shown in Figure~\ref{pow_2d_fig} for
simulations f2000 and f250 (solid lines). For reference we also show the
corresponding angular power spectra of the underlying density field (dotted
lines). These power spectra were constructed at full resolution without any
beam- or bandwidth smearing, since these are instrument-specific. The smallest
value of $\ell$ at which we calculate the power spectra corresponds to half of
our box size, since below that there is an unphysical damping due to the
finite simulation box size. The largest $\ell$ value shown corresponds to one
computational cell and thus is related to our spatial resolution.

The power spectrum at redshift $z=20.3$ is identical to that of the underlying
density field, since at that time there are only a few, small H~II regions. In
both reionization histories, evolution of the power spectrum proceeds by a
rise at small scales (large $\ell$), shifting slowly to larger scales as
reionization reaches the 50\% point, and followed by a decline of the power as
most of the universe gets ionized. The declining power spectrum keeps peaking
at approximately the same $\ell$ value as when it was at its maximum, but the
distribution becomes quite broad. Comparing the f2000 and f250 cases shows
that the f250 case peaks at somewhat smaller scales ($\ell\approx 5000$,
$\theta\approx4^\prime$) than the f2000 case ($\ell\approx 4000$,
$\theta\approx5^\prime$). The f250 distribution is also somewhat broader
with a slightly lower peak value.

This evolution of the power spectrum reflects the evolution of the H~II
regions.  Initially they are small, but dominate the fluctuations, resulting
in more fluctuation power at high $\ell$. The maximum then moves to lower
$\ell$ as the H~II regions grow with time. As we showed in Paper I this growth
levels off at a certain size ($\sim10$~Mpc for f2000) because once an ionized
bubble grows larger than that size it merges with many other bubbles to form
much larger ionized regions. This reflects a typical correlation length
between source clusters, convolved with the source efficiency (number of
ionizing photons produced per atom). The last is reflected in the f250 power
spectra peak being at lower scales (higher $\ell$). These very large bubbles
which result from the mergers of multiple smaller ones are rare, however, and
locally they partly retain the shapes and sizes of the original smaller
bubbles which merged (see e.g.  Figure~\ref{f2000_406}), so the typical
fluctuation size at the scales under consideration (sub-degree) is still
dictated by the typical size ($\sim10$~Mpc) of bubbles resulting from local
source clustering.

The power spectra magnitude grows strongly during the initial, very patchy
phase of reionization, in all cases reaching a maximum of $\sim10$~mK around
the time of 50\% ionization, in complete agreement with the results from the
rms fluctuations we discussed above in Sect.~\ref{rms_sect}. This is
substantially over the power contained in the density field, about a factor 2
for the f2000 case, and 1.5 for the f250 case. At late times ($z<12.3$ for
f2000) the power spectra are significantly depressed due to the small neutral
fraction remaining, with the peak value reaching only up to 1-2 mK. The
broadness of the peak indicates that the power is contained in ever larger
structures.

\begin{figure}
\begin{center}
\includegraphics[width=3.0in]{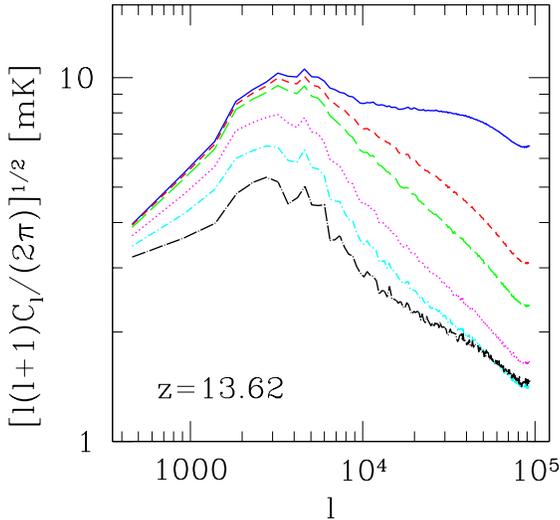}
\caption{Angular power spectra $[l(l+1)C_l/2\pi]^{1/2}$ at $z=13.62$ for 
  f2000\_406 and using bandwidths (top to bottom) 1 cell, 0.1 MHz, 0.2 MHz,
  0.5 MHz, 1 MHz, and 2 MHz. This illustrates the effect on the power spectrum
  of integration over a given bandwidth.
\label{pow_2d_diffbandw_fig}}
\end{center}
\end{figure}

We do not show how beam smoothing affects the power spectra as this
can be derived easily by multiplying the power spectrum with that of
the beam (Fig.~\ref{beams_ps}), and for detailed observational
predictions this should be done with the ``dirty beam'' of the
particular array under consideration. Both the Gaussian and
compensated Gaussian beam will introduce an effective cut-off at the
FWHM angle. Considering the multipole values for the peaks which we
find, this means that planned experiments may be just sufficient to
capture the expected peaks.  {Our results are qualitatively similar to
the analytical estimates in \citet{2004ApJ...613....1F,
2004ApJ...613...16F}. These authors also find that during reionization
a peak develops in the angular power spectrum, and as reionization
progresses it shifts to lower $\ell$. However, our peaks tend to be
more pronounced, and do not reach as low $\ell$'s as in the analytical
estimates.} This divergent behaviour reflects the differences in their
prescription for the growth of bubbles and our simulations.

The bandwidth smoothing can in principle be derived by applying the
appropriate window function to the 3D power spectra.  Qualitatively, bandwidth
smoothing takes away power at the high $\ell$ values since it will average of
small bubbles along the line of sight. To illustrate the effect we show in
Fig~\ref{pow_2d_diffbandw_fig} how bandwidth smoothing affects the power
spectrum at $z=13.62$ for the f2000\_406 case. The bandwidth varies from about
30 kHz (one cell) up to 2 MHz. Without smoothing the peak is at
$\ell\sim4000$, moving to larger scales for larger bandwidths. However, both
the magnitude and the position of the peak barely change for
$\Delta\nu\leq0.2$~MHz, staying close to the maximum value of $10.3$~mK. For
the maximum bandwidth considered, 2~MHz, the peak value drops by almost a
factor of 2, to 5.46 mK, and occurs at $\ell\sim3000$. As expected, even
modest smoothing washes out most of the small-scale power, but there is also
some decrease of power on the larger scales, since we sample larger scales at
which the fluctuations are lower.

\subsection{Beyond the power spectra: non-Gaussianity of the 21-cm signal}
\label{nongauss_sect}

\begin{figure*}
\begin{center}
\includegraphics[width=5in]{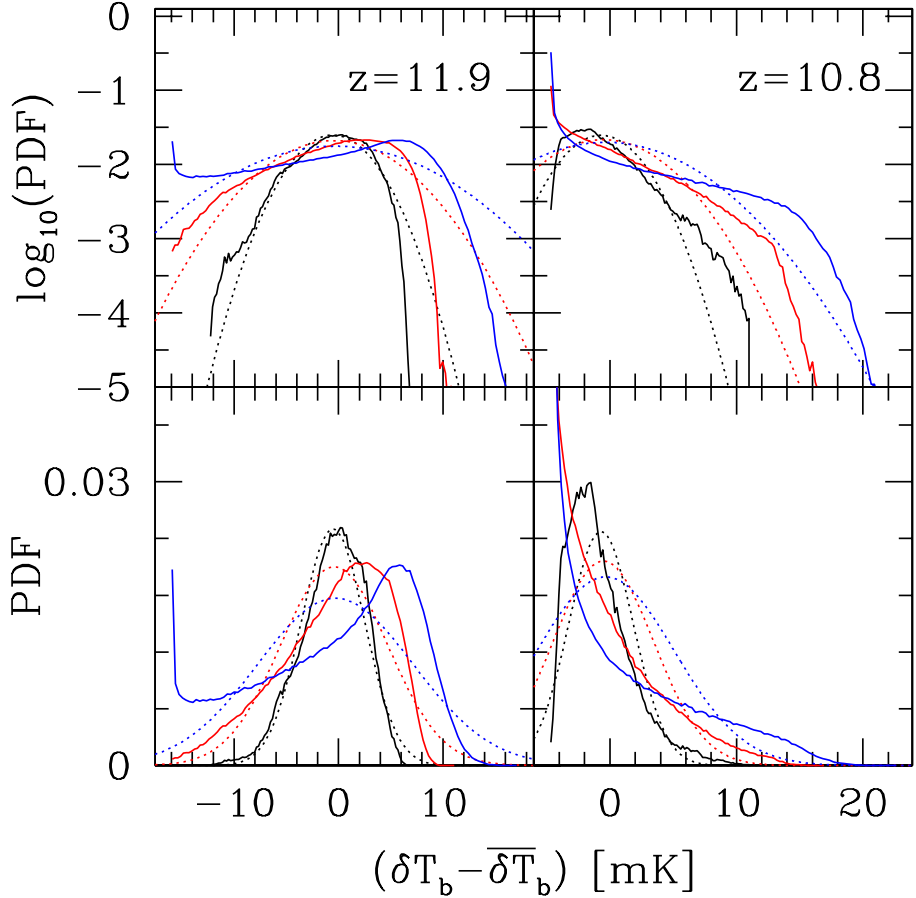}
\caption{Non-Gaussianity of the 21-cm signal: PDF distribution of the 21-cm 
  signal from simulation f250 for $z=11.9$ (left) and $z=10.8$ (right), shown
  both at logarithmic (top row) and linear (bottom row) scale. The
  PDF were derived for cubical regions of sizes $20\,h^{-1}Mpc$ (black solid),
  $10\,h^{-1}Mpc$ (red solid), and $5\,h^{-1}Mpc$ (blue solid). Also indicated
  are the Gaussian distributions with the same mean values and standard
  deviations (dotted, corresponding colours).
\label{nongauss}}
\end{center}
\end{figure*}

The power spectra discussed in the previous section provide important
information about the magnitude of the 21-cm fluctuations at different scales.
However, they fall well short of a full description of the statistics of the
21-cm fluctuations, and the corresponding implications for high-redshift
structure formation and reionization. Distributions with identical power
spectra could have completely different properties and statistics. In Paper~I
we showed quantitatively for the first time that the PDFs of both the ionized
gas fraction and the ionized gas mass are generally strongly non-Gaussian,
even more so than the underlying density distribution, which starts Gaussian,
but develops non-Gaussian features due to the formation of non-linear
structures. Here we extend our results further by studying the PDF of the
21-cm differential brightness temperature.

We calculate the PDFs using the same method as in Paper~I, in cubes of
5, 10 and 20$\,h^{-1}$~Mpc size {(approximately corresponding to
angular resolutions of $10^\prime$, $5^\prime$ and $2.^\prime 5$
respectively)} and at two different redshifts for simulation f250, the
first at the 50\% ionization ($z=11.9$) and the second during the late
stages of reionization ($z=10.8$). For the other simulations the
PDFs are similar at the same ionization levels (which correspond to
different redshifts due the different reionization histories). The
results are shown in Figure~\ref{nongauss} (solid lines), along with
the corresponding Gaussian distributions with the same means and
widths (dotted lines). These Gaussian distributions are the ones one
would construct on the basis of the power spectrum analysis. The PDFs
are expressed in terms of deviations from the mean differential
brightness temperature at that redshift.

These plots show that the PDFs of the 21-cm signal are highly non-Gaussian. 
Considering for instance the 5$\,h^{-1}$~Mpc case (blue lines) at $z=11.9$, 
there is a 5-10 times larger probability for finding holes (corresponding to 
ionized regions) deeper than $-10$~mK from the average signal (which is 16~mK
at that redshift) than expected from the Gaussian analysis. For the same 5~Mpc 
scale the peaks of 10~mK above the mean are also more frequent than the 
Gaussian statistics predicts, but higher peaks ($>12$~mK) are actually more 
rare. The same trends are seen at larger scales (10 and 20$\,h^{-1}$~Mpc), 
although to somewhat lesser extent.

At the later redshift of 10.8, at which time the mean signal is 5~mK, there
are very large over-abundances, by factors of up to 30, of holes of $-6$ to
$-2$~mK.  Even more interestingly, strong peaks of 10-20~mK above the mean
signal are up to 1-2 orders of magnitude more abundant than the corresponding
Gaussian predictions, and the distribution is even more skewed than at the
earlier time.  This over-abundance in the number of high peaks is roughly
independent of the spatial scale considered, but occurs at different
temperatures, from $\sim10$~mK for 20$\,h^{-1}$~Mpc regions, up to 16-18~mK
for 5$\,h^{-1}$~Mpc regions.

This shows that the even very late into reionization, when most of the
universe is ionized (for simulation f250 the mass ionized fraction is 84\% at
$z=10.8$), there is still considerable signal available in isolated
features. The results in Sect.~\ref{maxima} already hinted at this, and the
non-Gaussian statistics puts numbers to it. The same behaviour is seen in all
of our simulations, and should increase the chances of actually detecting the
redshifted 21-cm line, even if reionization was mostly complete rather early.

{We can compare our results to previously published reionization
  PDFs.  \citet{2003ApJ...596....1C} studied a 20$/h$~Mpc simulation
  volume and presented the PDF at the resolution of their pixel size,
  much below the resolution of planned observations and the sizes we
  are addressing here, making a comparison difficult.
  \citet{2004ApJ...613...16F} calculated PDFs from their
  semi-analytical excursion set model. The $z=13$ curve in their
  Fig.~4 can be roughly compared to our 5~Mpc curves at $z=11.9$ since
  both refer to a state of $\sim 50\%$ ionization seen at
  $\sim2.^\prime 5$. In order to facilitate the comparison,
  Fig.~\ref{nongauss} also shows our PDFs plotted on a linear
  scale. We see that our results differ substantially from theirs, not
  being Gaussian as in their outside-in and uniform reionization
  models, and not showing the sharp cut-off of their inside-out
  models.}

%%%%%%%%%%%%%%%%%%%%%%%%%%%%%%%%%%%%%%%%%%%%%%%%%%%%%%%%%%%%%%%%%%%%%%%%%

\section{Conclusions}
\label{summary_sect}

We constructed the evolution of the redshifted 21-cm emission for several 
reionization histories resulting from large-scale radiative transfer 
simulations in a volume of (100$h^{-1}$~Mpc)$^3$. This volume corresponds 
to $\sim1$ degree on the sky and $\sim5-10$~MHz in frequency, sufficient 
for capturing the scales relevant to reionization and its 21-cm signatures.
We studied the various aspects of these 21-cm signatures, from the 
properties and statistics of individual bright features, to sky radio maps, to
statistically-extracted signals. Our results can be summarized as follows. 

We do not find a very sharp global step in the 21-cm signal resulting from the
neutral gas being ionized and disappearing. Instead, the transitions we
observed are rather gradual, with a $\sim20$~mK decrement of the signal over
$\sim 20$~MHz. A similar results if obtained for the change in rms
fluctuations with frequency.  Assuming well-behaved foregrounds, such a
transition is well within the achievable instrument sensitivity and thus could
still be detectable, though not as easily as a sharper transition would have
been.

The large-scale geometry of reionization is well-resolved in 21-cm sky maps
even after beam smearing. Such maps would show structures in the neutral IGM,
as well as the shapes of the HII regions are the sources, providing a unique
window on early structure formation. We also showed that subtracting the
average signal from a large frequency band ($\sim$5MHz) should be efficient
way to remove an idealized smooth foreground while leaving the 21~cm
fluctuations essentially intact.
 
Analyzing high-resolution LOS spectra we find that redshift distortions are
considerable, and hence will influence the shapes of structures along the LOS,
and should be taken into account when constructing 3D (position/frequency)
observables from simulations. There is a large variation between different LOS
in terms of the structures encountered and the time when the epoch of overlap
is reached. If the foreground emission from galactic and extra-galactic
sources can be effectively removed, the fluctuations seen along the LOS would
provide detailed information about the underlying neutral density
field. However, the most easily detectable objects would be local ionized
bubbles, which are bordered by sharp transitions, where the 21~cm signal
changes by tens of mK over very small intervals in frequency. We also
presented for the first time simulated LOS spectra filtered with a realistic
beam and bandwidth. These show that although small details are smoothed over
and the signal variations are smaller than in the full-resolution spectra, the
main structures remain clearly visible and should be detectable.

Statistical measures, such as fluctuations and the two-dimensional power
spectra show that the fluctuations always reach a maximum when the gas is
$\sim 50$\% ionized by mass, and that the power is boosted considerably (by a
factor $\sim 2$) compared to the total density. At that maximum the power
spectra show a clear peak at an angular scale of $\ell\sim 3000-6000$,
depending on the particular reionization history. Earlier and later, the peaks
are broader, indicating that more scales, both larger and smaller are
involved. In order to be able to observe this peak the planned experiments
should have a resolution of $\sim 3^\prime$ or better.

The derived PDFs show that the 21~cm signal is strongly non-Gaussian at all
scales. In particular, at late times the chance of finding individual bright
features can be 10 or more times above the Gaussian prediction. At all times
the brightest point in our simulation volume is 20--60~mK for beam/bandwidth
smoothing typical for the planned LOFAR/PAST/GMRT-type observations. This is
insensitive to the particular reionization history. It is therefore likely
that even in the case of early reionization, individual bright peaks can be
observed at frequencies above 120~MHz.

Finally, we note that while we consider several specific reionization
scenarios here, many of our conclusions appear to be quite generic, e.g.\ the
significant boost of the 21-cm emission fluctuations due to reionization
patchiness, the fluctuations reaching maximum at $\sim50\%$ ionization by mass
and declining thereafter, and the relative insensitivity of the magnitude and
statistics of the brightest spots in the box. On the other hand, certain
important features vary significantly among scenarios - e.g. at which redshift
the fluctuations peak is reached. These depend on a number of still
highly-uncertain reionization parameters. In the models discussed here we vary
the most important of these parameters, the ionizing photon production
efficiency of the sources and the level of gas clumping at small scales. As we
discussed above, the ability to detect EOR signatures with any particular
instrument would depend strongly on when these features occur.

\section*{Acknowledgments} 
GM acknowledges support from the Royal Netherlands Academy of Art and Sciences.
This work was partially supported by NASA Astrophysical Theory Program grants
NAG5-10825 and NNG04GI77G to PRS. We thank P.~McDonald, C.~Hirata and S.~Sethi
for enlightening discussions, and TIARA for their hospitality and financial 
support. 

\bibliographystyle{mn2e} 
%\bibliography{../../refs}

\end{document}